\documentclass[12pt]{article}

\input epsf
\usepackage{amssymb}
\usepackage[dvips]{graphicx}
\usepackage{amsmath,amscd}

\def\dj{\hbox{d\kern-0.347em \vrule width 0.3em height 1.252ex depth
-1.21ex \kern 0.051em}}

\numberwithin{equation}{section}

\begin{document}

\setlength{\oddsidemargin}{0cm}
\setlength{\baselineskip}{7mm}


\thispagestyle{empty}
\setcounter{page}{0}

\begin{flushright}
CERN-PH-TH/083-2006   \\
LMU-ASC 36/06 \\
MPP-2006-51 \\
{\tt hep-th/0605113}
\end{flushright}

\vspace*{0.5cm}

\begin{center}
{\bf \Large Comments on noncommutative gravity}

\vspace*{1cm}

Luis \'Alvarez-Gaum\'e$\,^{\rm a,\,\,}$\footnote{\tt
Luis.Alvarez-Gaume@cern.ch},
Frank Meyer$\,^{\rm b,\,\,}$\footnote{\tt
meyerf@theorie.physik.uni-muenchen.de}
and Miguel A. V\'azquez-Mozo$^{\rm c,\,\, }$\footnote{\tt
Miguel.Vazquez-Mozo@cern.ch}

\end{center}

\vspace*{0.25cm}

\begin{quote}
   $^{\rm a}$ {\sl Theory Group, Physics Department CERN, CH-1211 Geneva
     23, Switzerland}


   $^{\rm b}$ {\sl Arnold Somerfeld Center for Theoretical Physics,
     Universit\"at M\"unchen, Theresienstra\ss e 37, D-80333 Munich,
     Germany {\rm and} Max-Planck-Institut f\"ur Physik, F\"ohriger
     Ring 6, D-80805 Munich, Germany.}

   $^{\rm c}$ {\sl Departamento de F\'{\i}sica Fundamental, Universidad
     de Salamanca, Plaza de la Merced s/n, E-37008 Salamanca, Spain
     {\rm and} Instituto Universitario de F\'{\i}sica Fundamental y
     Matem\'aticas (IUFFyM), Universidad de Salamanca, Salamanca,
     Spain}

\end{quote}

\vspace*{0.5cm}

\centerline{\bf \large Abstract}

\noindent
We study the possibility of obtaining noncommutative gravity dynamics
from string theory in the Seiberg-Witten limit. We find that the
resulting low-energy theory contains more interaction terms than those
proposed in noncommutative deformations of gravity. The r\^ole of
twisted diffeomorphisms in string theory is studied and it is found that
they are not standard physical symmetries. It is argued that this
might be the reason why twisted diffeomorphisms are not preserved by
string theory in the low energy limit. Twisted gauge transformations
are also discussed.

\newpage

\section{Introduction}

The construction of consistent noncommutative deformations of Einstein
gravity has been a subject of interest for some time (for an
incomplete list of references see
\cite{NCgravity,chamseddine,solutions}).  Following the standard
procedure to construct noncommutative deformations of gauge and scalar
theories
\cite{reviews}, noncommutative versions of the Einstein-Hilbert action
have been obtained by replacing the ordinary product by the
noncommutative Moyal product
\begin{eqnarray}
f(x)\star g(x)=f(x)e^{{i\over 2}\theta^{\mu\nu}\overleftarrow 
{\partial}_{\mu}
\overrightarrow{\partial}_{\nu}}g(x).
\label{moyal}
\end{eqnarray}
There are many ways in which these deformations have been implemented.
Generically, noncommutative deformations of gravity lead to a
complexification of the metric as well as of the local Lorentz
invariance of the theory which is no longer SO(1,3) but U(1,3) or
larger \cite{chamseddine}. This results in theories with hermitian
metrics reminiscent of those studied long ago by Einstein and
Straus \cite{einstein_straus}.  These  theories
are known to contain ghost states
\cite{damour_et_al} (for some proposal to overcome these problems see
\cite{solutions}).

An unsatisfactory feature of many of the approaches developed so far
is that they are physically {\it ad hoc}, since the deformation of
general relativity is not based on any general dynamical
principle. Recently, however, a new approach has been proposed in
which a deformation of the Einstein-Hilbert action is constructed based
on a deformation of the group of diffeomorphisms \cite 
{munich1,munich2} (see
\cite{meyer_review} for a review). The idea behind it is to replace the
diffeomorphism invariance of general relativity by a twisted version
of this symmetry \cite{twist}, i.e. deforming the Hopf algebra
structure of the universal enveloping algebra of the Lie algebra of
vector fields by twisting the coproduct in the appropriate way (see
also \cite{others}). Roughly speaking this amounts to keeping the
action of the diffeomorphisms on the physical ``primary'' fields
unchanged while deforming the Leibniz rule when taking the action of
diffeomorphisms on the product of two fields. This change have the
effect that diffeomorphisms now act ``covariantly'' on the
star-product of two fields.

Once the deformation of the diffeomorphisms has been introduced a
gravity action can be written with the requirement of invariance
under the action of the twisted transformations. As with other
proposals for noncommutative gravity, the deformed Einstein-Hilbert
action can be written in terms of star-products.  Nevertheless, unlike
other proposals, this construction has the obvious advantage of being
based on an underlying symmetry principle.

One may wonder whether string theory can reproduce the
deformed gravity action proposed in \cite{munich1,munich2} in some
limit. It is known that the strict Seiberg-Witten
\cite{seiberg_witten} limit results in a complete decoupling of  
gravity. This
is because in order to retain nontrivial gauge dynamics on
the brane one has to scale the closed string coupling constant to zero
in the low-energy limit. As a result, closed string states decouple at
low energies and the resulting theory is not coupled to gravity.

By looking at the next-to-leading order in the Seiberg-Witten limit it
is possible to study the dynamics of closed strings in the presence of
a constant $B$-field. In this paper we will study the gravitational
action induced by the bosonic string theory on a space-filling D-brane
with a constant magnetic field in the low energy limit. We find that
the induced terms for the interaction vertex of three gravitons on the
brane contain terms which cannot be derived from an action expressed
solely in terms of star-products and therefore are not accounted for
in the noncommutative gravity action proposed in
\cite{munich1,munich2}. Moreover, these new terms scale in the same
way in the Seiberg-Witten limit as the ones associated with
star-products and we have found no physical argument that allows their
consistent elimination to reproduce an induced gravity action that can
be expressed in terms of star-products alone.

We will see that this inability to reproduce a gravity action
invariant under deformed diffeomorphisms can be traced back to the
fact that the star-product is not ``inert'' under the action of
space-time diffeomorphisms.  As it will be shown below, the twisted
Leibniz rule can be interpreted as resulting from applying the
ordinary Leibniz rule but taking into account the transformation of
the star-product.  Hence the invariance of the deformed gravity
action under twisted diffeomorphisms cannot be considered a physical
symmetry in the standard sense, since the transformation involves not
only the physical fields but also the star-products. Thus,
string theory cannot provide the deformed diffeomorphisms at low
energies since these transformations do not yield a physical symmetry
of the theory.

The present paper is organized as follows: in Section \ref{induced} we
will review the construction of the induced gravity action on D-branes
in the standard case. After that, in Section \ref{NCaction}, we will
summarize the relevant aspects of the construction of the
noncommutative gravity action presented in \cite{munich1,munich2}. We  
move on
in Section \ref{NCinduced} to study the gravity action induced on
the brane in the presence of a $B$-field to the next-to-leading order
in the Seiberg-Witten limit. As advertised, we will find extra terms
not present in the field theory construction of the
noncommutative action. Finally, in Section \ref{explanation}
we will try to understand in physical terms the mismatch between the
string and field theory constructions, before summarizing our results
in Section \ref{conclusions}. In order to make the presentation more
self-contained, Appendix A contains a summary of results and
definitions concerning Hopf algebras.

\section{Brane induced gravity from string theory}
\label{induced}

We begin by computing the gravitational action induced on the brane in
the absence of a $B$-field. For simplicity we work here with the
bosonic string in the presence of a space-filling D-brane.  The
analysis can be extended to superstrings and/or lower-dimensional
D-branes \cite{corley_lowe_ramgoolam}.

We carry out the computation by evaluating the correlation
functions of three graviton vertex operators on the disk. These
correlation functions induce three-graviton interaction terms in
the gravitational action on the brane. It is important to stress at
this point that we are computing the gravitational action {\sl induced}
on the brane. It is well known that the graviton amplitudes on the
disk contain divergences associated with the coincidence limit of two
or more vertex operator insertions that are usually handled
by considering the different factorization limits including
the contributions of Riemann surfaces without boundaries
\cite{tseytlin}.  In our case, however, we eliminate these divergences
by adding appropriate counterterms in the induced action on the brane.
Then the divergences in the string amplitudes will be absorbed in a
renormalization of the coefficients multiplying the different terms in
the induced gravitational action. From  this point of view  the
origin of the terms in the induced action is similar to the way in
which $\lambda\phi^{3}$ and $\lambda\phi^{4}$ interaction terms are
induced in the Yukawa theory.

We have to calculate the disk correlation function of three graviton
vertex operators \cite{string_books,dhoker_phong}
\begin{eqnarray}
V_{p}={g_{s}\over \alpha'}\varepsilon_{\mu\nu}(p)\int d^{2}z \partial  
X^{\mu}
\overline{\partial}X^{\nu}e^{ip\cdot X}(z,\overline{z}),
\hspace*{1cm} \varepsilon_{[\mu\nu]}(p)=0=
\eta^{\mu\nu}\varepsilon_{\mu\nu}(p),
\end{eqnarray}
where the symmetric-traceless polarization tensor
has to satisfy the transversality conditions
$p^{\mu}\varepsilon_{\mu\nu}(p)$ and the momentum has to be on-shell,
$p^{2}=0$, for the vertex operator to be a primary field with
conformal weight (1,1). Naively, the coupling between three on-shell
gravitons (or gauge fields) vanishes for kinematical reasons. In our
case we remain on-shell while continuing the momenta to complex value,
in such a way that the resulting amplitudes give a nonzero result from
which the induced couplings can be read off.

The relevant quantity to evaluate is the correlation function
of three graviton vertex operators
\begin{eqnarray}
\langle V_{p_{1}}V_{p_{2}}V_{p_{3}}\rangle_{D}&=&{g_{s}^{2}\over
(\alpha')^{3}}
\int_{D}\prod_{i=1}^{3}d^{2}z_{i}  \langle\!\langle\prod_{k=1}^{3}
\partial X^{\mu}
\overline{\partial}X^{\nu}e^{ip\cdot X}(z_{k},\overline{z}_{k})
\rangle\!\rangle_{D},
\end{eqnarray}
where $\langle\!\langle\ldots\rangle\!\rangle_{D}$ indicates the
correlation function of the corresponding operator on the disk. In
order to carry out the calculation, it is very convenient to
rewrite the polarization tensors as $\varepsilon_{\mu\nu}\equiv
\zeta_{\mu}\overline{\zeta}_{\nu}$ so that the graviton
vertex operator can be expressed as
\begin{eqnarray}
V_{p}={g_{s}}\left.\int
d^{2}z\,e^{i\mathcal{P}\cdot X}\right|_{\zeta\overline{\zeta}}
\end{eqnarray}
where by the subscript we indicate that only the part linear in
$\zeta_{\mu}\overline{\zeta}_{\nu}$ has to be kept, and
$\mathcal{P}_{\mu}$ is defined by
\begin{eqnarray}
\mathcal{P}_{\mu}\equiv p_{\mu}-{i\over\sqrt{\alpha'}}
\zeta_{\mu}\partial-{i\over\sqrt{\alpha'}}\overline{\zeta}_{\mu}
\overline{\partial}.
\end{eqnarray}

Proceeding in this way the correlation function of three gravitons can
be written in a way similar to the three
tachyons amplitude with the momenta $p_{\mu}$ replaced by $\mathcal{P} 
_{\mu}$,
\begin{eqnarray}
\langle V_{p_{1}}V_{p_{2}}V_{p_{3}}\rangle_{D} &=&{g_{s}^{2}\over
(\alpha')^{13}}
(2\pi)^{26}\delta(p_{1}+p_{2}+p_{3})
\int_{D}\left.\prod_{i=1}^{3}d^{2}z_{i}
\prod_{k<\ell}^{3}e^{-\mathcal{P}_{k}\cdot\mathcal{P}_{\ell}G(z_{k},z_ 
{\ell})}
\right|_{(\zeta\overline{\zeta})^{3}},
\label{amplitude_3g}
\end{eqnarray}
where the delta function and the powers of $\alpha'$ arise from the
integration over the zero modes of $X^{\mu}(z,\overline{z})$ and
$G(z,w)$ is the propagator of the Laplacian on the disk
\begin{eqnarray}
G(z,w)=-\alpha'\Big(\log|z-w|-\log|z-\overline{w}|\Big).
\end{eqnarray}

If we use now the momentum conservation together with the on-shell
condition for the momenta, the exponent inside the integral in
Eq. (\ref{amplitude_3g}) simplifies to
\begin{eqnarray}
\sum_{k<\ell}^{3}\mathcal{P}_{k}\cdot\mathcal{P}_{\ell}G(z_{k},z_ 
{\ell})&=&
-{i\over\sqrt{\alpha'}}
\sum_{k\neq \ell}^{3}\left[(p_{k}\cdot\zeta_{\ell})\partial_{\ell}+
(p_{k}\cdot\overline{\zeta}_{\ell})
\overline{\partial}_{\ell}\right]G(z_{k},z_{\ell})
\nonumber \\
&-&{1\over 2\alpha'}\sum_{k\neq\ell}^{3}
\left[(\zeta_{k}\cdot\zeta_{\ell})\partial_{k}\partial_{\ell}
+(\overline{\zeta}_{k}\cdot\overline{\zeta}_{\ell})\overline{\partial} 
_{k}
\overline{\partial}_{\ell}\right]G(z_{k},z_{\ell}) \\
&-&{1\over \alpha'}
\sum_{i\neq j}^{3}(\zeta_{i}\cdot\overline{\zeta}_{j})
\partial_{i}\overline{\partial}_{j}
G(z_{i},z_{j}). \nonumber
\end{eqnarray}
In order to obtain the terms in the effective action we have to expand
the exponential in Eq.  (\ref{amplitude_3g}),
$\exp\left[-\sum_{k<\ell}^{3}
\mathcal{P}_{k}\cdot\mathcal{P}_{\ell}G(z_{k},z_{\ell})\right]$
keeping the terms linear in $\zeta_{\mu}\overline{\zeta}_{\nu}$. Doing
this we obtain terms with two, four and six momenta that induce
interactions in the action that are associated respectively with terms
linear, quadratic and cubic in the Riemann tensor \cite{tseytlin} (the
term without derivatives is associated to a ``cosmological'' term
proportional to $\sqrt{-g}$).

The only terms that contributes to the Einstein-Hilbert term are those
with two momenta given by
\begin{eqnarray}
\left.e^{-\sum_{k<\ell}^{3}
\mathcal{P}_{k}\cdot\mathcal{P}_{\ell}G(z_{k},z_{\ell})}
\right|_{(\zeta\overline{\zeta})^{3}}
&=&
-{1\over(\alpha')^{3}}
\sum_{i,\ell=1}^{3}\left[\sum_{(j,a,b)=1}^{3}\sum_{(m,c,d)=1}^{3}
(p_{i}\cdot\zeta_{j})(p_{\ell}\cdot\overline{\zeta}_{m})(\zeta_{a}\cdot
\zeta_{b})(\overline{\zeta}_{c}\cdot\overline{\zeta}_{d})\right.
\nonumber \\
& & \times\,\, \partial_{j}G_{\rm reg}(z_{i},z_{j})
\overline{\partial}_{m}G_{\rm reg}(z_{\ell},z_{m})
\partial_{a}\partial_{b}G_{\rm reg}(z_{a},z_{b})
\overline{\partial}_{c}\overline{\partial}_{d}
G_{\rm reg}(z_{c},z_{d}) \nonumber \\
&+&
\sum_{(j,a,c)=1}^{3}\sum_{[m,b,d]=1}^{3}(p_{i}\cdot\zeta_{j})
(p_{\ell}\cdot\overline{\zeta}_{m})(\zeta_{a}\cdot\overline{\zeta}_{b})
(\zeta_{c}\cdot\overline{\zeta}_{d}) \nonumber \\
& &  \times\,\,\partial_{j}G_{\rm reg}(z_{i},z_{j})\overline{\partial} 
_{m}
G_{\rm reg}(z_{\ell},z_{m})\partial_{a}\overline{\partial}_{b}
G_{\rm reg}(z_{a},z_{b})\partial_{c}\overline{\partial}_{d}
G_{\rm reg}(z_{c},z_{d}) \nonumber \\
&+&
\sum_{(a,j,m)=1}^{3}\sum_{\{a,b,c\}=1}^{3}(p_{i}\cdot\zeta_{j})
(p_{\ell}\cdot{\zeta}_{m})(\zeta_{a}\cdot\overline{\zeta}_{b})
(\overline{\zeta}_{c}\cdot\overline{\zeta}_{d}) \\
& &  \times\,\,\partial_{j}G_{\rm reg}(z_{i},z_{j}){\partial}_{m}
G_{\rm reg}(z_{\ell},z_{m})\partial_{a}\overline{\partial}_{b}
G_{\rm reg}(z_{a},z_{b})\overline{\partial}_{c}
\overline{\partial}_{d}G_{\rm reg}(z_{c},z_{d}) \nonumber \\
&+&
\sum_{(a,j,m)=1}^{3}\sum_{\{m,b,c\}=1}^{3}(p_{i}\cdot\overline{\zeta}_ 
{j})
(p_{\ell}\cdot\overline{\zeta}_{m})(\zeta_{a}\cdot{\zeta}_{b})
(\overline{\zeta}_{c}\cdot{\zeta}_{d}) \nonumber \\
& &  \times\,\,\overline{\partial}_{j}G_{\rm reg}(z_{i},z_{j})
\overline{\partial}_{m}
G_{\rm reg}(z_{\ell},z_{m})\partial_{a}{\partial}_{b}
G_{\rm reg}(z_{a},z_{b})\overline{\partial}_{c}
{\partial}_{d}G_{\rm reg}(z_{c},z_{d}) \Bigg],\nonumber
\end{eqnarray}
where we have used the notation $(a,b,c)$ to indicate a sum where all
the three indices are different and $b<c$, $[a,b,c]$ to denote that
all three indices are different and in addition $b>d$ and $\{a,b,c\}$
that the three indices are different without any further constraint.
We have also indicated that the propagators have to be regularized in
the coincidence limit for the expression to be finite.

A calculation of the corresponding integrals gives the following
expression for the term with two momenta \cite{tseytlin}
\begin{eqnarray}
\langle V_{1}V_{2}V_{3}\rangle_{p^{2}}
&=&\mathcal{A}\Big[(p_{3}\cdot\varepsilon_{1}
\cdot p_{3})(\varepsilon_{2}\cdot\varepsilon_{3})+(p_{1}\cdot 
\varepsilon_{3}
\cdot p_{1})(\varepsilon_{1}\cdot\varepsilon_{2})+(p_{2}\cdot 
\varepsilon_{1}
\cdot p_{2})(\varepsilon_{2}\cdot\varepsilon_{3}) \nonumber \\
&+&2(p_{1}\cdot\varepsilon_{3}\cdot\varepsilon_{2}\cdot\varepsilon_{1}
\cdot p_{2})
+2(p_{2}\cdot\varepsilon_{1}\cdot\varepsilon_{3}\cdot\varepsilon_{2}
\cdot p_{3})
+2(p_{3}\cdot\varepsilon_{2}\cdot\varepsilon_{1}\cdot\varepsilon_{3}
\cdot p_{1})\Big],
\hspace*{1cm}
\label{3p}
\end{eqnarray}
where the coefficient $\mathcal{A}$ is given by\footnote{To avoid  
cumbersome
expressions we have factored out explicitly the momentum conservation
delta function $(\alpha')^{-13}(2\pi)^{26}\delta(p_{1}+p_{2}+p_{3})$.}
\begin{eqnarray}
\mathcal{A}=-{g_{s}^{2}\over(\alpha')^{3}}
\int_{D}\prod_{i=1}^{3}d^{2}z_{i}\Big|(\partial_{2}G_{12}
-\partial_{2}G_{32})\partial_{1}
\partial_{3}G_{13}\Big|^{2}.
\label{A3p}
\end{eqnarray}
This integral has divergences associated with the coincidence limits
of two or more insertions which, by conformal invariance, are related
to the factorization limits containing closed string amplitudes. These
divergences can be regularized in a way compatible with both conformal
invariance and target-space general covariance \cite{tseytlin}. In
our case, as explained above, these divergences will be reabsorbed in
a renormalization of the coupling constant.

In constructing the induced graviton interactions on the brane we have
to be careful in normalizing the graviton field properly. In
particular, because the disk amplitude with two graviton vertex
operators is nonvanishing \cite{tseytlin} and scales as $g_{s}$, we
have to reabsorb powers of the string coupling constant in the
definition of the graviton field in such a way that the quadratic term
in the effective action is independent of $g_{s}$. Therefore we take
the following correspondence between the graviton wave function
$h_{\mu\nu}(x)$ and the polarization tensor $\varepsilon_{\mu\nu}(p)$
\begin{eqnarray}
g_{s}^{1\over 2}(\alpha')^{7}\int {d^{26}p\over (2\pi)^{26}}
\varepsilon_{\mu\nu}(p)e^{ip\cdot x}\longrightarrow h_{\mu\nu}(x).
\label{rescaling}
\end{eqnarray}
With this normalization, it is easy to see that the perturbative
expansion can be written as an expansion not in powers of the closed
string coupling, $g_{s}$, but of the so-called open string coupling
constant, $g_{o}=g_{s}^{1\over 2}$. Then, the gravitational
constant $\kappa$ scales as
\begin{eqnarray}
\kappa \sim g_{s}^{1\over 2}(\alpha')^{6}=g_{o}(\alpha')^{6}.
\end{eqnarray}
Notice that this scaling with the string coupling constant is
different from the one that emerges in the string low energy effective
action, $\kappa \sim g_{s}(\alpha')^{6}$. The reason behind is of course
that, strictly speaking, here we are not computing the low energy
effective action but the gravity action induced on the brane at low
energies.

Taking into account the previous discussion, the corresponding term in
the effective action induced by the part of the amplitude represented  
by (\ref{3p})
can be computed from
\begin{eqnarray}
\Delta S_{2}=(\alpha')^{26}\int {d^{26}p_{1}\over (2\pi)^{26}}
{d^{26}p_{2}\over (2\pi)^{26}}
{d^{26}p_{3}\over (2\pi)^{26}}
\,(2\pi)^{26}
\delta(p_{1}+p_{2}+p_{3})\,
\langle V_{p_{1}}V_{p_{2}}V_{p_{3}}\rangle_{p^{2}},
\label{extraS}
\end{eqnarray}
leading to the following induced term in the Lagrangian
\begin{eqnarray}
\Delta\mathcal{L}_{2}=-2\kappa\Big(2h^{\sigma\mu}\partial_{\mu}h^ 
{\alpha\beta}
\partial_{\beta}h_{\alpha\sigma}
+h^{\mu\sigma}h_{\alpha\beta}\partial_{\sigma}\partial_{\mu}
h^{\alpha\beta}\Big).
\label{term2p}
\end{eqnarray}
This term can be obtained from the Einstein-Hilbert term
$\mathcal{L}_{\rm EH}={1\over 2\kappa^{2}} \sqrt{-g}R$ in the weak
field expansion near Minkowski space-time,
$g_{\mu\nu}=\eta_{\mu\nu}+ 2\kappa h_{\mu\nu}$.

The terms with four and six momenta induce contributions
to the action containing higher powers of the curvature
tensor.  With four momenta we find the
following tensor structure
\begin{eqnarray}
\langle V_{1}V_{2}V_{3}\rangle_{p^{4}}
&=&\mathcal{B}\Big[(p_{1}\cdot\varepsilon_{2}\cdot
\varepsilon_{3}\cdot p_{1})(p_{2}\cdot\varepsilon_{1}\cdot p_{3})+
(p_{2}\cdot\varepsilon_{3}\cdot
\varepsilon_{1}\cdot p_{2})(p_{3}\cdot\varepsilon_{2}\cdot p_{1})  
\nonumber \\
&+&(p_{3}\cdot\varepsilon_{1}\cdot
\varepsilon_{2}\cdot p_{3})(p_{1}\cdot\varepsilon_{3}\cdot p_{2})\Big]
\end{eqnarray}
where the coefficient $\mathcal{B}$ is given by
\begin{eqnarray}
\mathcal{B}&=&-2{g_{s}^{2}\over(\alpha')^{3}}\mbox{Re}\,\,
\int_{D}\prod_{i=1}^{3}d^{2}z_{i}\Big|(\partial_{2}G_{12}-\partial_{2} 
G_{32})\Big|^{2}
\Big[(\partial_{1}G_{12}-\partial_{1}G_{31})(\partial_{2}G_{13}- 
\partial_{3}G_{23})
\overline{\partial}_{2}\overline{\partial}_{3}G_{23} \nonumber \\
&+&(\overline{\partial}_{1}G_{12}-\overline{\partial_{1}}G_{31})
(\partial_{2}G_{13}-\partial_{3}G_{23})
\partial_{2}\overline{\partial}_{3}G_{23}\Big]
\label{B}
\end{eqnarray}
and we assume that the divergences arising from the coincidence limits
have been properly regularized.  As in the previous case, this
amplitude induces a term in the effective action that can be computed
as in Eq. (\ref{extraS}) with the result
\begin{eqnarray}
\Delta\mathcal{L}_{4}&=&-8a_{4}\alpha'
\kappa h^{\mu\nu}\partial_{\mu}\partial_{\sigma}h^{\alpha\beta}
\partial_{\alpha}\partial_{\beta}h_{\nu}^{\,\,\,\,\sigma},
\label{term4p}
\end{eqnarray}
where $a_{4}$ is a numerical constant.
As explained in
\cite{tseytlin}, a term of this type arises from the weak field  
expansion
of a piece in the effective action quadratic in the curvature tensor,
${\alpha'a_{4}\over 2\kappa^{2}}\sqrt{-g}
R_{\mu\nu,\sigma\lambda}R^{\mu\nu,\sigma\lambda}$.

Finally we analyze the term with six momenta. From the calculation of
the disk amplitude we get
\begin{eqnarray}
\langle V_{1}V_{2}V_{3}\rangle_{p^{6}}
&=&\mathcal{C}(p_{1}\cdot\varepsilon_{2}\cdot p_{1})
(p_{2}\cdot\varepsilon_{3}\cdot p_{2})(p_{3}\cdot\varepsilon_{1}\cdot  
p_{3}),
\end{eqnarray}
with
\begin{eqnarray}
\mathcal{C}=-{g_{s}^{2}\over(\alpha')^{3}}
\int_{D}\prod_{i=1}^{3}d^{2}z_{i}\Big|(\partial_{2}G_{12}-\partial_{2} 
G_{32})\Big|^{2}
\Big|(\partial_{3}G_{23}-\partial_{3}G_{13})\Big|^{2}
\Big|(\partial_{1}G_{31}-\partial_{1}G_{21})\Big|^{2}.
\label{C}
\end{eqnarray}
Following the same procedure as above, it can be seen that this part
of the three-graviton amplitude induces the following term in the
effective action
\begin{eqnarray}
\Delta\mathcal{L}_{6}
=8a_{6}(\alpha')^{2}\kappa\partial_{\mu}\partial_{\nu}h^{\alpha\beta}
\partial_{\alpha}
\partial_{\sigma}h^{\lambda\nu}\partial_{\lambda}\partial_{\beta}
h^{\mu\sigma}.
\label{term6p}
\end{eqnarray}
Again $a_{6}$ is a new dimensionless coupling.  This piece of the  
induced
action is the leading term of
${a_{6}(\alpha')^{2}\over 2\kappa^{2}}\sqrt{-g}
R^{\mu\nu}_{\,\,\,\,\,\,\,\,\alpha\beta}
R^{\alpha\beta}_{\,\,\,\,\,\,\,\,\sigma\lambda}
R^{\sigma\lambda}_{\,\,\,\,\,\,\,\,\mu\nu}$ in the expansion around
flat space-time.

Putting together the previous results (\ref{term2p}), (\ref{term4p})
and (\ref{term6p}) we conclude that open strings induce a three-point
graviton action which reproduces the weak-field expansion of the
induced action
\begin{eqnarray}
S_{\rm ind}={1\over 2\kappa^{2}}\int
d^{26}x\sqrt{-g}\Big[{a_{0}\over \alpha'}+R+a_{4}\alpha'
R_{\mu\nu,\sigma\lambda}R^{\mu\nu,\sigma\lambda}+a_{6}(\alpha')^{2}
R^{\mu\nu}_{\,\,\,\,\,\,\,\,\alpha\beta}
R^{\alpha\beta}_{\,\,\,\,\,\,\,\,\sigma\lambda}
R^{\sigma\lambda}_{\,\,\,\,\,\,\,\,\mu\nu}+\ldots\Big].
\end{eqnarray}
Terms containing higher powers of the curvature tensor do not
contribute to the three-graviton amplitude. In the spirit of induced
gravity, the divergences contained in the integrals in Eqs. (\ref{B})
and (\ref{C}) are absorbed into the renormalized couplings $a_{0}$,
$\kappa$, $a_{4}$ and $a_{6}$.

\section{Noncommutative gravity}
\label{NCaction}

Einstein's General Relativity, including the cosmological constant term,
can be derived by requiring general covariance and that the
action contains only up to two derivatives of the metric. In the case of
noncommutative gravity a similar approach has been advanced in
\cite{munich1,munich2} where local diffeomorphisms are twisted. In the
following we review some basic aspects of this approach and study
the corrections to the standard gravitational action in the weak field
expansion around flat space-time. The extra couplings between
gravitons depending on the noncommutativity parameter $\theta^{\mu\nu}$
are the ones we will try to obtain from string theory in Section
\ref{NCinduced}.

\subsection{Deformed diffeomorphisms}

As already explained above, the starting point in the approach of
\cite{munich1,munich2} is the deformation of diffeomorphisms by
twisting. In General Relativity, infinitesimal
diffeomorphisms are generated by vector fields
$\xi(x)=\xi^{\sigma}\partial_{\sigma}$. Their action on the physical
tensor fields $T^{(p)}_{(q)}(x)$ of type $(p,q)$ is given by
\begin{eqnarray}
T\,{'}^{(p)}_{(q)}(x+\xi)=T^{(p)}_{(q)}(x),
\end{eqnarray}
where $T\,{'}^{(p)}_{(q)}(x)$ represents the transformed tensor field.
By expanding to linear order in $\xi$ the transformation of
$T^{(p)}_{(q)}(x)$ can be written in terms or its Lie derivative as
\begin{eqnarray}
\delta_{\xi} T^{(p)}_{(q)}(x)\equiv
T^{(p)}_{(q)}(x+\xi)-T^{(p)}_{(q)}(x)=-\mathcal{L}_{\xi} T^{(p)}_{(q)} 
(x),
\label{diffeos}
\end{eqnarray}
or in components
\begin{eqnarray}
\delta_{\xi}T^{\mu_{1}\ldots\mu_{p}}_{\nu_{1}\ldots\nu_{q}}(x)&=&
-\xi^{\alpha}(x)\partial_{\alpha}
T^{\mu_{1}\ldots\mu_{p}}_{\nu_{1}\ldots\nu_{q}}(x)+
\partial_{\alpha}\xi^{\mu_{1}}(x)T^{\alpha\ldots\mu_{p}}_{\nu_{1}
\ldots\nu_{p}}(x)
+\ldots+
\partial_{\alpha}\xi^{\mu_{n}}(x)T^{\mu_{1}\ldots\alpha}_{\nu_{1}
\ldots\nu_{q}}(x)
\nonumber \\
&-&\partial_{\nu_{1}}\xi^{\beta}(x)
T^{\mu_{1}\ldots\mu_{p}}_{\beta\ldots\nu_{q}}(x)
-\ldots-
\partial_{\nu_{q}}\xi^{\beta}(x)T^{\mu_{1}
\ldots\mu_{p}}_{\nu_{1}\ldots\beta}(x).
\end{eqnarray}
In addition, the infinitesimal diffeomorphism generated by $\xi$ acts
on the product of two tensor fields $T^{\,\,\,(p_{1})}_{1\,\,(q_{1})} 
(x)$,
$T^{\,\,\,(p_{2})}_{2\,\,(q_{2})}(x)$ via the Leibniz rule
\begin{eqnarray}
\delta_{\xi}\left[T^{\,\,\,(p_{1})}_{1\,\,(q_{1})}(x)
T^{\,\,\,(p_{2})}_{2\,\,(q_{2})}(x)
\right]=\delta_{\xi}T^{\,\,\,(p_{1})}_{1\,\,(q_{1})}(x)\,
T^{\,\,\,(p_{2})}_{2\,\,(q_{2})}(x)+T^{\,\,\,(p_{1})}_{1\,\,(q_{1})} 
(x)\,\,
\delta_{\xi}
T^{\,\,\,(p_{2})}_{2\,\,(q_{2})}(x).
\label{leibnizT}
\end{eqnarray}

In a theory deformed by replacing standard products by the Moyal
star-products (\ref{moyal}) the application of the Leibniz rule
(\ref{leibnizT}) shows that the product of two tensor fields does not
transform covariantly, i.e. unlike in the case analyzed above
$T^{\,\,(p_{1})}_{1\,\,(q_{1})}(x)\star
T^{\,\,(p_{2})}_{2\,\,(q_{2})}(x)$ does not transform as a tensor of
type $(p_{1}+p_{2},q_{1}+q_{2})$. This means that a noncommutative
gravity theory based on replacing standard (commutative) products by
star-products will fail to be invariant under diffeomorphisms.

This fact, however, does not necessarily mean that there are no other
transformations leaving invariant the action of the deformed
theory and that, in the limit $\theta^{\mu\nu}\rightarrow 0$, reduce to
the standard diffeomorphism invariance. There are in principle two
possible ways to find the deformed transformations. The first one is
to deform the action of infinitesimal diffeomorphisms (\ref{diffeos})
on the fields in such a way that, using the Leibniz rule, the deformed
action remains invariant. This is analogous to how gauge invariance gets
deformed to star-gauge invariance in gauge theories on noncommutative  
spaces.

A second alternative is to keep the transformations (\ref{diffeos})
intact but to deform the way it acts on products of fields. In
\cite{munich1,munich2} a twist of the standard diffeomorphisms has
been constructed to achieve precisely this. The idea behind it is
the realization that the universal enveloping algebra of
vector fields has a Hopf algebra structure (some basic facts about
Hopf algebras, as well as the notation used here, are summarized in
Appendix A). In particular, for an infinitesimal diffeomorphism generated by
$\xi\neq \mathbf{1}$, the coproduct $\Delta(\xi)$ can be defined to
be\footnote{In the case of infinitesimal diffeomorphisms, the identity
$\mathbf{1}$ correspond to the diffeomorphism generated by a vanishing
vector field}
\begin{eqnarray}
\Delta(\xi)=\xi\otimes \mathbf{1}+\mathbf{1}\otimes\xi,
\label{untwisted_coproduct}
\end{eqnarray}
while $\Delta(\mathbf{1})=\mathbf{1}\otimes\mathbf{1}$.

In fact, the choice of the coproduct determines the way the algebra of
diffeomorphisms acts on the products of fields. For a given vector
field $\xi$, the action on the product of two tensor fields
$T^{\,\,\,(p_{1})}_{1\,\,(q_{1})}(x)$,
$T^{\,\,\,(p_{2})}_{2\,\,(q_{2})}(x)$ is given by [cf. Eq.
(\ref{compatibility_action})]
\begin{eqnarray}
\delta_{\xi}[T^{\,\,\,(p_{1})}_{1\,\,(q_{1})}(x)
T^{\,\,\,(p_{2})}_{2\,\,(q_{2})}(x)]\equiv \mu\left\{ 
\Delta(\xi) \left[
T^{\,\,\,(p_{1})}_{1\,\,(q_{1})}(x)\otimes
T^{\,\,\,(p_{2})}_{2\,\,(q_{2})}(x)\right]\right\}.
\end{eqnarray}
Using the coproduct (\ref{untwisted_coproduct}) the
standard Leibniz rule (\ref{leibnizT}) is retrieved.

Equation (\ref{untwisted_coproduct}) is not the only possible choice for
a coproduct to define a Hopf algebra structure in the algebra of
infinitesimal diffeomorphisms. In particular it is possible to introduce
the twist operator
\begin{eqnarray}
\mathcal{F}=e^{-{i\over 2}\theta^{\mu\nu}\partial_{\mu}\otimes\partial_ 
{\nu}},
\label{twist_operator}
\end{eqnarray}
in terms of which a new twisted coproduct is defined as
\begin{eqnarray}
\Delta(\xi)_{\mathcal{F}}=\mathcal{F}\left(
\xi\otimes \mathbf{1}+\mathbf{1}\otimes\xi\right)\mathcal{F}^{-1}.
\label{twisted_coproduct}
\end{eqnarray}
The twist operator (\ref{twist_operator}) enters also in the definition
of the star-product of two fields
\begin{eqnarray}
T^{\,\,\,(p_{1})}_{1\,\,(q_{1})}(x)\star
T^{\,\,\,(p_{2})}_{2\,\,(q_{2})}(x) =
\mu\left\{\mathcal{F}^{-1}\left[T^{\,\,\,(p_{1})}_{1\,\,(q_{1})}(x)\otimes
T^{\,\,\,(p_{2})}_{2\,\,(q_{2})}(x)\right]\right\}\equiv \mu_{\star} 
\left[
T^{\,\,\,(p_{1})}_{1\,\,(q_{1})}(x)\otimes
T^{\,\,\,(p_{2})}_{2\,\,(q_{2})}(x)\right],
\end{eqnarray}
which is just a more sophisticated way of writing Eq. (\ref{moyal}).

It is important to stress that the twist does not change the action
of infinitesimal diffeomorphisms on the fields, that it is still given
by (\ref{diffeos}). The effect of the twist is to change the
coproduct and consequently the action of diffeomorphisms on the product.
By using Hadamard's formula
\begin{eqnarray}
e^{A}Be^{-A}=\sum_{n=0}^{\infty}{1\over n!}[\,\underbrace{A,[A,\ldots 
[A}_{n},B]
\ldots]],
\label{hadamard}
\end{eqnarray}
the twisted coproduct (\ref{twisted_coproduct}) can be written in powers
of $\theta^{\mu\nu}$ as
\begin{eqnarray}
\Delta(\xi)_{\mathcal{F}}&=&\xi\otimes\mathbf{1}+\mathbf{1}\otimes\xi
\\
&+& \sum_{n=1}^{\infty}{(-i/2)^{n}\over n!}\theta^{\mu_{1}\nu_{1}}
\ldots\theta^{\mu_{n}\nu_{n}}\Big\{[\partial_{\mu_{1}},[\partial_{\mu_ 
{2}},
\ldots[\partial_{\mu_{n}},\xi]\ldots]]
\otimes\partial_{\nu_{1}}\partial_{\nu_{2}}
\ldots\partial_{\nu_{n}} \nonumber \\
& & +\,\,\partial_{\mu_{1}}\partial_{\mu_{2}}\ldots\partial_{\mu_{n}}
\otimes[\partial_{\nu_{1}},[\partial_{\nu_{2}},\ldots[\partial_{\nu_ 
{n}},\xi]
\ldots]]\Big\}. \nonumber
\end{eqnarray}
This twisted coproduct defines the action of the Lie algebra of
vector fields on the star-product of two fields as
\begin{eqnarray}
& & \delta_{\xi}\Big[T^{\,\,\,(p_{1})}_{1\,\,(q_{1})}(x)\star
T^{\,\,\,(p_{2})}_{2\,\,(q_{2})}(x)\Big]\equiv \mu_{\star} 
\left\{
\Delta(\xi)_{\mathcal{F}} \Big[T^{\,\,\,(p_{1})}_{1\,\,(q_{1})} 
(x)\otimes
T^{\,\,\,(p_{2})}_{2\,\,(q_{2})}(x)\Big]\right\} \nonumber \\
& & \hspace*{1cm} =\,\,\delta_{\xi}
T^{\,\,\,(p_{1})}_{1\,\,(q_{1})}(x)\star
T^{\,\,\,(p_{2})}_{2\,\,(q_{2})}(x)+ T^{\,\,\,(p_{1})}_{1\,\,(q_{1})} 
(x)\star
\delta_{\xi}T^{\,\,\,(p_{2})}_{2\,\,(q_{2})}(x) \nonumber\\
& & \hspace*{1cm}
+\,\,\sum_{n=1}^{\infty}{(-i/2)\over n!}\theta^{\mu_{1}\nu_{1}}
\ldots\theta^{\mu_{n}\nu_{n}}\Big\{[\partial_{\mu_{1}}[\partial_{\mu_ 
{2}},
\ldots[\partial_{\mu_{n}},\delta_{\xi}]\ldots]]
T^{\,\,\,(p_{1})}_{1\,\,(q_{1})}(x)\star\partial_{\nu_{1}}\partial_ 
{\nu_{2}}
\ldots\partial_{\nu_{n}}T^{\,\,\,(p_{2})}_{2\,\,(q_{2})}(x)
\nonumber \\
& & \hspace*{1cm}+\,\,
\partial_{\mu_{1}}\partial_{\mu_{2}}\ldots\partial_{\mu_{n}}
T^{\,\,\,(p_{1})}_{1\,\,(q_{1})}(x)\star
[\partial_{\nu_{1}},[\partial_{\nu_{2}},\ldots[\partial_{\nu_{n}}, 
\delta_{\xi}]
\ldots]]T^{\,\,\,(p_{2})}_{2\,\,(q_{2})}(x)\Big\}.
\label{twisted_leibniz}
\end{eqnarray}
The interesting thing of this choice of the coproduct is that
unlike the standard Leibniz rule, Eq. (\ref{twisted_leibniz})
guarantees that the star-product of two tensor fields of type
$(p_{1},q_{1})$ and $(p_{2},q_{2})$ is a tensor of type
$(p_{1}+p_{2},q_{1}+q_{2})$. Therefore the star-product transforms
``covariantly'' with respect to twisted diffeomorphisms.

\subsection{The deformed gravity action and its weak field expansion}

Once the deformation of diffeomorphisms has been carried out, it is
possible to proceed with the construction of the deformed gravity
action. In \cite{munich1,munich2} a gravity action is presented and
shown to be invariant under deformed diffeomorphisms.  In
$D$-dimensions this action reads\footnote{ Following the notation of
Refs. \cite{munich1,munich2} we denote by $\hat{G}_{\mu\nu}$ the
deformed metric and by $g_{\mu\nu}$ its commutative limit
$\theta^{\mu\nu}\rightarrow 0$. For other quantities, we denote the
deformed ones with a hat.}
\begin{eqnarray}
S_{\star{\rm EH}}={1\over 2\kappa^{2}}
\int d^{D}x \Big[(\det{}_{\star}\,e_{\mu}^{\,\,\,\,a})
\hat{G}^{\mu\nu}\star R_{\mu\nu}+\mbox{c.c.}\Big].
\label{EH-deformed}
\end{eqnarray}
Here $\hat{G}^{\mu\nu}$ denotes the star-inverse of
$\hat{G}_{\mu\nu}$, i.e.  $ \hat{G}_{\sigma\nu}\star
\hat{G}^{\mu\sigma}=\delta^{\mu}_{\,\,\,\,\nu}$. In terms of the
vielbein $e_{\mu}^{\,\,\,\,a}$ the deformed metric is given by
\begin{eqnarray}
\hat{G}_{\mu\nu}={1\over 2}
\Big(e_{\mu}^{\,\,\,\,a}\star
e_{\nu}^{\,\,\,\,b}+e_{\nu}^{\,\,\,\,a}\star e_{\mu}^{\,\,\,\,b}
\Big)\eta_{ab}
=g_{\mu\nu}-{1\over 8}\theta^{\alpha_{1}\beta_{1}}\theta^{\alpha_{2} 
\beta_{2}}
(\partial_{\alpha_{1}}\partial_{\alpha_{2}}e_{\mu}^{\,\,\,\,a})
(\partial_{\beta_{1}}\partial_{\beta_{2}}e_{\nu}^{\,\,\,\,b})+\ldots,
\label{defmetric}
\end{eqnarray}
and the star-determinant of the vielbein is defined by
\begin{eqnarray}
\det{}_{\star}e_{\mu}^{\,\,\,\,a}
={1\over D!}\epsilon^{\mu_{1}\ldots\mu_{D}}
\epsilon_{a_{1}\ldots a_{D}}e_{\mu_{1}}^{\,\,\,\,a_{1}}\star\ldots\star
e_{\mu_{D}}^{\,\,\,\,a_{D}}.
\end{eqnarray}
The other ingredient of the deformed Einstein-Hilbert action
(\ref{EH-deformed}), the Ricci tensor, is defined in terms of the
deformed Riemann tensor
\begin{eqnarray}
\hat{R}_{\mu\nu,\sigma}^{\,\,\,\,\,\,\,\,\,\,\,\,\,\,\lambda}\equiv  
\partial_{\nu}
\hat{\Gamma}_{\mu\sigma}^{\,\,\,\,\,\,\,\,\lambda}-\partial_{\mu}
\hat{\Gamma}_{\nu\sigma}^{\,\,\,\,\,\,\,\,\lambda}+
\hat{\Gamma}_{\nu\sigma}^{\,\,\,\,\,\,\,\,\alpha}
\star\hat{\Gamma}_{\mu\alpha}^{\,\,\,\,\,\,\,\,\lambda}-
\hat{\Gamma}_{\mu\sigma}^{\,\,\,\,\,\,\,\,\alpha}
\star\hat{\Gamma}_{\nu\alpha}^{\,\,\,\,\,\,\,\,\lambda}
\end{eqnarray}
by $\hat{R}_{\mu\nu}\equiv
\hat{R}_{\mu\sigma,\nu}^{\,\,\,\,\,\,\,\,\,\,\,\,\sigma}$. In turn,
the deformed Christoffel symbols
$\hat{\Gamma}_{\mu\nu}^{\,\,\,\,\,\,\,\,\sigma}$ can be computed from  
the
metric $\hat{G}_{\mu\nu}$ and its star-inverse by
\begin{eqnarray}
\hat{\Gamma}_{\mu\nu}^{\,\,\,\,\,\,\,\,\sigma}
={1\over 2}\Big(\partial_{\mu}G_{\nu\alpha}
+\partial_{\nu}G_{\mu\alpha}
-\partial_{\alpha}G_{\mu\nu}\Big)\star \hat{G}^{\alpha\sigma}.
\end{eqnarray}
As shown in \cite{munich1,munich2} the deformed action
(\ref{EH-deformed}) is invariant under the deformed algebra of
diffeomorphisms, $\delta_{\xi}S_{\star EH}=0$, acting on the fields
in the way defined in the previous subsection.

Our main goal is to decide whether the deformed Einstein-Hilbert
action (\ref{EH-deformed}) follows somehow from string theory, and the
way in which we will do it is by comparing (\ref{EH-deformed}) to the
terms in the gravitational action induced on the brane in the presence
of a constant $B$-field. Since the string theory calculation proceeds
by evaluating scattering amplitudes of gravitons, it is convenient
to write the deformed gravity action in a weak-field expansion around
flat space-time by writing
\begin{eqnarray}
g_{\mu\nu}=\eta_{\mu\nu}+2\kappa h_{\mu\nu}.
\label{weak_field_expansion}
\end{eqnarray}
In the case of the vielbeins the weak field expansion is implemented
by\footnote{In the Euclidean case this relation can be obtained by
noticing that in matrix notation
$\mathbf{g}=\mathbf{e}\mathbf{e}^{T}$.  In addition, the real matrix
$\mathbf{e}$ can be written as the product of an orthogonal and a
symmetric matrix, $\mathbf{e}=\mathbf{S}\mathbf{O}$. Therefore
$\mathbf{g}=\mathbf{S}^{2}$ and $\mathbf{S}$ is given by the square
root of the metric. Then in the weak field expansion
$\mathbf{S}=\mathbf{1}+ {1\over 2}(2\kappa\mathbf{h})$ and,
fixing the gauge freedom to $\mathbf{O}=\mathbf{1}$, we find
that $\mathbf{S}=\mathbf{e}\equiv \mathbf{1}+(2\kappa)\boldsymbol 
{\tau}$.}
\begin{eqnarray}
e_{\mu}^{\,\,a}=\delta_{\mu}^{\,\,a}+2\kappa\tau_{\mu}^{\,\,a},
\label{weak_e}
\end{eqnarray}
where $\tau_{\mu}^{\,\,a}$ is related to the graviton field by
\begin{eqnarray}
\tau_{\mu}^{\,\,a}={1\over 2}\eta^{ab}h_{\mu\nu}\delta^{\nu}_{\,\,a}.
\end{eqnarray}
As above, we  fix the gauge invariance by taking the
transverse-traceless gauge, $h^{\mu}_{\,\,\mu}=0$,
$\partial_{\mu}h^{\mu\nu}=0$.

Moreover, in order to eventually compare with the string theory
calculation presented in Section \ref{NCinduced} in
the weak field expansion (\ref{weak_field_expansion}) we will
consider only the leading terms in the expansion in the
noncommutativity parameter $\theta^{\mu\nu}$.  Using (\ref{defmetric})
and (\ref{weak_e}) we find for the deformed metric
\begin{eqnarray}
\hat{G}_{\mu\nu}=\eta_{\mu\nu}
+2\kappa h_{\mu\nu}-{\kappa^{2}\over 32}\theta^{\alpha_{1}\beta_{1}}
\theta^{\alpha_{2}\beta_{2}}
\partial_{\alpha_{1}}\partial_{\alpha_{2}}h_{\mu}^{\,\,\,\,\sigma}
\partial_{\beta_{1}\beta_{2}}h_{\nu\sigma}+\mathcal{O}(\kappa^{3}, 
\theta^{4}),
\end{eqnarray}
whereas for its inverse there is also a linear term in
$\theta^{\mu\nu}$
\begin{eqnarray}
\hat{G}^{\mu\nu}&=&
\eta^{\mu\nu}-2\kappa h^{\mu\nu}
+\kappa^{2}\Big[4h^{\mu\sigma}h_{\sigma}^{\,\,\,\,\nu}
-2i\theta^{\alpha\beta}\partial_{\alpha}h^{\mu\sigma}
\partial_{\beta}h^{\mu}_{\,\,\,\,\sigma}
\nonumber \\
&-&{3\over 8}\theta^{\alpha_{1}\beta_{1}}
\theta^{\alpha_{2}\beta_{2}}(\partial_{\alpha_{1}}
\partial_{\alpha_{2}}h^{\mu\sigma})
(\partial_{\beta_{1}}\partial_{\beta_{2}}h^{\nu}_{\,\,\,\,\sigma})
\Big]+\mathcal{O}(\kappa^{3},\theta^{3}).
\end{eqnarray}
In the case of the star-determinant of the vielbein
$\det{}_{\star}e_{\mu}^{\,\,\,\,a}$, as it also happens in the case in
Einstein gravity, the only term contributing to the three-graviton
amplitude is the leading one, $\det{}_{\star}e_{\mu}^{\,\,\,\,a}=1+
\mathcal{O}(\kappa,\theta)$.

With these ingredients we can proceed to compute the leading
$\theta$-dependent terms in the three-graviton amplitude. Because the action
(\ref{EH-deformed}) is real the first correction to the
Einstein-Hilbert action has to contain two powers of the
noncommutativity parameter. Hence, for dimensional reasons, we find
that the first nontrivial contribution to the three-graviton vertex
contains two powers of $\theta$ and six derivatives. Collecting all
the terms of this type we find that the leading correction to the
three-graviton vertex is
\begin{eqnarray*}
\Delta_{\theta}\mathcal{L}_{\rm EH}=\left.{1\over 2\kappa^{2}}\Big[
\hat{G}^{\mu\nu(2)}R_{\nu\mu}
+g^{\mu\nu}\hat{R}^{(2)}_{\nu\mu}+\hat{G}^{\mu\nu(1)}
\hat{R}_{\nu\mu}^{(1)}\Big]\right|_{h^{3}}+\mbox{total derivatives,}
\end{eqnarray*}
where by the superindex $(k)$ we denote the terms with $k$ powers of
$\theta^{\mu\nu}$ and the subindex $h^{3}$ indicates that we are
keeping only the terms with three graviton fields.  Using the
expressions given above for the different terms and after a long but
straightforward calculation we find the sought term to be
\begin{eqnarray}
\Delta_{\theta}\mathcal{L}_{\rm EH}
&=&{1\over 2}\kappa\theta^{\nu\gamma}\theta^{\eta\rho}
\Big(2h^{\sigma\mu}\partial_{\mu}\partial_{\nu}\partial_{\eta}h^ 
{\alpha\beta}
\partial_{\beta}\partial_{\gamma}\partial_{\rho}
h_{\alpha\sigma}+\partial_{\gamma}\partial_{\rho}
h^{\mu\sigma}h_{\alpha\beta}\partial_{\sigma}\partial_{\mu}
\partial_{\nu}\partial_{\eta}
h^{\alpha\beta}\Big) \nonumber \\
&+&\mbox{terms vanishing on-shell.}
\label{comparison}
\end{eqnarray}

In our analysis we are going to ignore those terms in the action that
are zero by applying the equations of motion. The reason is that
ultimately we want to compare this result with the induced gravity
action obtained from the string theory amplitudes.  Since string
theory only allows the computation of on-shell scattering amplitudes,
any term in the effective action vanishing on-shell cannot be
accounted for.  In order to compare with later results, we can rewrite
this term in momentum space as
\begin{eqnarray}
\Delta_{\theta}\mathcal{L}_{\rm EH}&=&
-{\kappa\over 12}(p_{2\,\mu}\theta^{\mu\nu}
p_{3\,\nu})^{2}
\Big[2p_{3\,\sigma}p_{2\,\lambda}h^{\sigma\beta}(p_{1})h_{\beta\eta} 
(p_{2})
h^{\eta\lambda}(p_{3})
+h^{\sigma\lambda}(p_{1})p_{2\,\alpha}p_{2\,\beta}h_{\sigma\lambda}(p_ 
{2})
h^{\alpha\beta}(p_{3})\Big] \nonumber\\ &+& \mathcal{O}(p_{i}\cdot
p_{j}).
\label{weak_hNC}
\end{eqnarray}
We notice that the term in brackets has the same tensor structure as
the standard Einstein-Hilbert terms for three gravitons,
Eq. (\ref{3p}). The reason is simple. Since we are expanding around
flat space, the kinetic term in the graviton action $\mathcal{L}_{\rm
kin}= {1\over 2}h_{\mu\nu}(-\nabla^{2}) h^{\mu\nu}$ is invariant under
Lorentz transformations.  In the absence of noncommutativity this
symmetry would be preserved by all the terms in the weak-field
expansion of the Einstein-Hilbert Lagrangian, unlike in our case where
the presence of $\theta^{\mu\nu}$ breaks that symmetry. However, in
the terms computed we find that the only source of Lorentz violation is
the overall factor $(p_{2\,\mu}\theta^{\mu\nu} p_{3\,\nu})^{2}$ while  
the
rest of the expression, not containing any power of $\theta^{\mu\nu}$,
has to preserve Lorentz invariance. As we know, there is only one  
term with
this property that contains two momenta and three graviton
fields, and this is precisely the three-graviton vertex of the  Einstein
action given in Eq. (\ref{3p}).

\section{Induced noncommutative gravity and the Seiberg-Witten limit}
\label{NCinduced}

We now turn to the question of whether the term (\ref{comparison}) can
be obtained in some low energy limit of string theory. As
in Section \ref{induced} we restrict our attention to the case of the
bosonic string on a D25-brane. Graviton interactions on
lower-dimensional branes have been considered in \cite{ardalan_et_al}.
Applications of the Seiberg-Witten map to the study of induced gravity
on noncommutative spaces have been studied in \cite{inducedNCgravity}.

In the Seiberg-Witten limit \cite{seiberg_witten} the low-energy limit
$\alpha'\rightarrow 0$ is taken while keeping fixed the open string
metric $G_{\mu\nu}$, the noncommutativity parameter $\theta^{\mu\nu}$
and the gauge coupling constant $g_{\rm YM}$. This limit can be
implemented by introducing a control parameter $\epsilon\rightarrow 0$
an scaling $\alpha'$, the closed string metric $g_{\mu\nu}$ and the
closed string coupling constant $g_{s}$ according to
\begin{eqnarray}
\epsilon^{1\over 2}\alpha', \hspace*{1cm}
\epsilon\,g_{\mu\nu}\hspace*{0.5cm}{\mbox{and}} \hspace*{0.5cm}
\epsilon^{1\over 2}g_{s},
\end{eqnarray}
where we have assumed that the $B$-field has maximal rank $r=24$.

 From the scaling of $g_{s}$ we see how closed string states decouple
in the Seiberg-Witten limit. The resulting low-energy field
theory is not coupled to gravity and no gravity action can be obtained
in this low-energy limit. Nevertheless, the gravitational couplings
can be studied by considering terms which are subleading in the
Seiberg-Witten limit, i.e. terms in the action which scale with
positive powers of $\epsilon$ in the limit $\epsilon\rightarrow
0$. Since we are interested in reproducing the $\theta$-dependent
terms in the three-graviton interaction vertex, the relevant string
amplitude to compute is the disk with three graviton vertex operators
insertions.  The most important change with respect to the standard
calculation presented in \cite{tseytlin} is that now, due to the
presence of the $B$-field, the disk propagator has the form
\begin{eqnarray}
\langle X^{\mu}(z,\overline{z})X^{\nu}(w,\overline{w})\rangle_{D}
&=&-\alpha'\left[
{1\over\sqrt{\epsilon}}g^{\mu\nu}\Big(\log|z-w|
-\log|z-\overline{w}|\Big)\right.
\nonumber \\
&+&\left.\sqrt{\epsilon}(-\Theta^{2})^{\mu\nu}\log|z-\overline{w}|^ 
{2}
+\Theta^{\mu\nu}\log\left({z-\overline{w}\over \overline{z}-w}\right) 
\right],
\end{eqnarray}
where we have introduced the dimensionless noncommutativity parameter
$\Theta^{\mu\nu}={1\over 2\pi\alpha'}\theta^{\mu\nu}$.

Once the disk propagator is known, the three-graviton amplitude can be
computed using the same techniques applied in the calculation
presented in Section \ref{induced}, namely
\begin{eqnarray}
\langle V_{p_{1}}V_{p_{2}}V_{p_{3}}\rangle_{D} &=&\epsilon^{-{11\over  
2}}\,
{g_{s}^{2}\over (\alpha')^{13}}
(2\pi)^{26}\delta(p_{1}+p_{2}+p_{3}) \nonumber \\
&\times &
\int_{D}\prod_{i=1}^{3}d^{2}z_{i}
\exp\left\{-\sum_{k<\ell}^{3}\Big[{1\over \sqrt{\epsilon}}
\mathcal{P}_{k}\cdot\mathcal{P}_{\ell}G(z_{k},z_{\ell})
\right. \label{amplitude_theta} \\
&+&\left.\left.\!\!\!\!
\sqrt{\epsilon}\mathcal{P}_{k}\bullet\mathcal{P}_{\ell}H(z_{k},z_ 
{\ell})+
\mathcal{P}_{k}\wedge\mathcal{P}_{\ell}K(z_{k},z_{\ell})\Big]
-{1\over 2}\sqrt{\epsilon}
\sum_{k=1}^{3}\mathcal{P}_{k}\bullet\mathcal{P}_{k}H(z_{k},z_{k})\right\}
\right|_{(\zeta\overline{\zeta})^{3}}, \nonumber
\end{eqnarray}
where $G(z,w)$ is the propagator on the disk,
\begin{eqnarray}
H(z,w)&=&-\alpha'\log|z-\overline{w}|^{2} \nonumber \\
K(z,w)&=& -\alpha'\log\left({z-\overline{w}\over \overline{z}-w}\right)
\end{eqnarray}
and we have introduced the notation
\begin{eqnarray}
a\bullet b\equiv a_{\mu}(-\Theta^{2})^{\mu\nu}b_{\nu}, \hspace*{1cm}
a\wedge b\equiv a_{\mu}\Theta^{\mu\nu}b_{\nu}.
\end{eqnarray}
It is also important to keep in mind that the $\bullet$- and
$\wedge$-products contain all the dependence on the noncommutativity
parameter. Notice also that we have factored out all powers of $ 
\epsilon$
so that we have a good control on the Seiberg-Witten limit. This  
means that
$\mathcal{P}_{\mu}$ is now given by
\begin{eqnarray}
\mathcal{P}_{\mu}\equiv p_{\mu}-{i\over \epsilon^{1\over 4}\sqrt 
{\alpha'}}
\zeta_{\mu}\partial-{i\over\epsilon^{1\over 4}\sqrt{\alpha'}}\overline 
{\zeta}_{\mu}
\overline{\partial}.
\end{eqnarray}

In order to reproduce the first $\theta$-dependent correction in the
noncommutative gravity action (\ref{EH-deformed}) we need terms
containing two $\theta$'s and six momenta. There are multiple ways of
obtaining this term from Eq. (\ref{amplitude_theta}). In particular,
bringing down all the terms of the form $(p_{i}\wedge
p_{j})(p_{k}\wedge p_{\ell}) K(z_{i},z_{j})K(z_{k},z_{\ell})$ from the
exponent, we find the following contribution to the
amplitude\footnote{At this stage one should not worry about the
overall negative power of $\epsilon$, since this is only due to the
normalization chosen for the correlation function. As we see below the
properly normalized induced term in the effective action is suppressed
by a positive power of $\epsilon$.}
\begin{eqnarray}
& & \hspace*{-1cm}
\epsilon^{-{11\over 2}}\,{g_{s}^{2}\over (\alpha')^{13}} \left.
(2\pi)^{26}\delta(p_{1}+p_{2}+p_{3})(p_{1}\wedge p_{2})^{2}
\int_{D}\prod_{i=1}^{3}d^{2}z_{i}\,K(z_{1},z_{2})^{2}
\prod_{k<\ell}^{3}e^{-{1\over \sqrt{\epsilon}}
\mathcal{P}_{k}\cdot\mathcal{P}_{\ell}G(z_{k},z_{\ell})}
\right|_{(\zeta\overline{\zeta})^{3}},\hspace*{0.5cm}
\end{eqnarray}
where in the exponential we have to keep the terms with two momenta
and three polarization tensors. Since all the dependence on the
noncommutativity parameter is already factored out, we can use our
results of Section \ref{induced} to write the amplitude as
\begin{eqnarray}
\widehat{\mathcal{A}}\,(p_{1}\wedge p_{2})^{2}\Big[(p_{3}\cdot 
\varepsilon_{1}
\cdot p_{3})(\varepsilon_{2}\cdot\varepsilon_{3})+
2(p_{1}\cdot\varepsilon_{3}\cdot\varepsilon_{2}\cdot\varepsilon_{1} 
\cdot p_{2})
+\mbox{permutations}\Big],
\label{wanted}
\end{eqnarray}
with
\begin{eqnarray}
\widehat{\mathcal{A}}=-\epsilon^{-{5\over 2}}
{\,g_{s}^{2}\over(\alpha')^{3}}
\int_{D}\prod_{i=1}^{3}d^{2}z_{i}\,K(z_{1},z_{2})^{2}
\Big|(\partial_{2}G_{12}-\partial_{2}G_{32})\partial_{1}
\partial_{3}G_{13}\Big|^{2}
\end{eqnarray}
and we have not included explicitly the overall factor implementing
momentum conservation, $(\sqrt{\epsilon}\alpha')^{-13}(2\pi)^{26}
\delta(p_{1}+p_{2}+p_{3})$.
We proceed now as in Section \ref{induced} by rescaling the graviton
wave function according to (\ref{rescaling}). Using Eq. (\ref{extraS})
the corresponding induced term in the action can be computed to be
\begin{eqnarray}
\epsilon^{21\over 2}
b_{2}\kappa\theta^{\nu\gamma}\theta^{\eta\rho}
\Big(2h^{\sigma\mu}\partial_{\mu}\partial_{\nu}\partial_{\eta}h^ 
{\alpha\beta}
\partial_{\beta}\partial_{\gamma}\partial_{\rho}
h_{\alpha\sigma}+\partial_{\gamma}\partial_{\rho}
h^{\mu\sigma}h_{\alpha\beta}\partial_{\sigma}\partial_{\mu}
\partial_{\nu}\partial_{\eta}
h^{\alpha\beta}\Big),
\end{eqnarray}
with $b_{2}$ the dimensionless coupling.  This term is precisely of
the form found in Eq. (\ref{weak_hNC}) for the first
$\theta$-dependent correction to the Einstein-Hilbert action. We
notice that, as expected, the induced term is suppressed by an overall
positive power of $\epsilon$.

Unfortunately together with the wanted term (\ref{wanted}) string
theory gives us a whole plethora of other terms which are not found in
the noncommutative gravity action (\ref{EH-deformed}) and that,
moreover, scale the same way as (\ref{wanted}) in the Seiberg-Witten
limit. In particular we can look at the term obtained from
(\ref{amplitude_theta}) by taking down all the terms proportional to
$p_{k}\bullet p_{\ell}$ ($k\neq \ell$) from the exponential
\begin{eqnarray}
& & \epsilon^{-5}\,{g_{s}^{2}}
(2\pi)^{26}\delta(p_{1}+p_{2}+p_{3})(p_{1}\bullet p_{2}+p_{1}\bullet
p_{3}+p_{2}\bullet p_{3})
\nonumber \\
& &\left.\hspace*{2cm} \times \,\,\,\int_{D}\prod_{i=1}^{3}d^{2}z_{i} 
\,H(z_{1},z_{2})
\prod_{k<\ell}^{3}e^{-{1\over \sqrt{\epsilon}}
\mathcal{P}_{k}\cdot\mathcal{P}_{\ell}G(z_{k},z_{\ell})}
\right|_{(\zeta\overline{\zeta})^{3}}.
\end{eqnarray}
Since we have factored out two momenta and two powers of the
noncommutative parameters we need to keep the terms in the expansion
of the exponential that contain three polarization tensors and four
momenta. Again we can go back to Section \ref{induced} to write
\begin{eqnarray}
& &\widehat{\mathcal{B}}\,\,(p_{1}\bullet p_{2}+p_{1}\bullet p_{3}
+p_{2}\bullet p_{3})
\Big[(p_{1}\cdot\varepsilon_{2}\cdot
\varepsilon_{3}\cdot p_{1})(p_{2}\cdot\varepsilon_{1}\cdot p_{3})+
(p_{2}\cdot\varepsilon_{3}\cdot
\varepsilon_{1}\cdot p_{2})(p_{3}\cdot\varepsilon_{2}\cdot p_{1})  
\nonumber \\
& &\hspace*{1cm}+\,\,(p_{3}\cdot\varepsilon_{1}\cdot
\varepsilon_{2}\cdot p_{3})(p_{1}\cdot\varepsilon_{3}\cdot p_{2})\Big],
\end{eqnarray}
where now after factoring out
$(\alpha'\sqrt{\epsilon})^{-13}(2\pi)^{26}\delta(p_{1}+p_{2}+p_{3})$
we have
\begin{eqnarray}
\widehat{\mathcal{B}}&=&-\epsilon^{-{5\over 2}}
{2g_{s}^{2}\over(\alpha')^{3}}
\int_{D}\prod_{i=1}^{3}d^{2}z_{i}\Big|\partial_{2}G_{12}
-\partial_{2}G_{32}\Big|^{2}
H(z_{1},z_{2}) \Big[(\partial_{1}G_{12}-\partial_{1}G_{31})
\\
&\times& (\partial_{2}G_{13}-\partial_{3}G_{23})
\overline{\partial}_{2}\overline{\partial}_{3}G_{23}
+(\overline{\partial}_{1}G_{12}-\overline{\partial_{1}}G_{31})
(\partial_{2}G_{13}-\partial_{3}G_{23})
\partial_{2}\overline{\partial}_{3}G_{23}+\mbox{c.c.}\Big]. \nonumber
\end{eqnarray}
This amplitude induces a term in the action is not present in the
weak-field expansion of the deformed action (\ref{EH-deformed}) and that
furthermore cannot be written in terms of star-products. The problem
however lies in that the new induced term has exactly the same scaling
in the Seiberg-Witten limit, $\epsilon^{21\over 2}$, as the one in
Eq. (\ref{wanted}), so both terms in the effective action are equally
important in the low energy limit.

Actually this is not the end of the story, since there are many other
terms which contain other $\theta$-dependent couplings and that are of
the same order in the low-energy limit. These, for example, include in
principle terms proportional to
\begin{eqnarray}
(p_{i}\wedge \varepsilon_{k}\wedge p_{i})\equiv p_{i\,\mu}\Theta^{\mu 
\nu}
\varepsilon_{k\,\nu\sigma}\Theta^{\sigma\lambda}p_{i\,\lambda}
\hspace*{0.5cm}
\mbox{and} \hspace*{0.5cm} (p_{i}\bullet \varepsilon_{k}\cdot p_{i})
\equiv p_{i\,\mu}(-\Theta^{2})^{\mu\nu}\varepsilon_{k\,\nu\sigma}p_{i} 
^{\sigma}.
\label{extra-terms}
\end{eqnarray}
Again, these terms scale in the Seiberg-Witten limit with the same
power of $\epsilon$ as the others we have kept\footnote{In the
philosophy of induced gravity it is possible to reabsorb powers of
$\epsilon$ in the graviton wavefunction in order that, for example,
the two-graviton induced term scales like $\epsilon^{0}$. This changes
the overall scaling of the three-graviton interaction terms containing
two powers of $\theta^{\mu\nu}$ from $\epsilon^{21\over 2}$ to
$\epsilon^{3}$ but does not change the relative scaling between the
different terms.}. As in the case of the other term discussed above the
induced interactions in the action cannot be expressed in terms of
star-products of the fields.

The result of our calculation is that in taking the low-energy limit
of gravitons interacting with open strings there are terms in the
induced effective action that cannot be written only in terms of
star-products. In the end this should not be a surprise, since this is
also known to happen when studying the coupling of open string states
to closed strings \cite{garousi,star-trek}.  In particular, couplings
of the type (\ref{extra-terms}) are found in Ref. \cite{garousi} when
computing the coupling of gauge fields to closed string tachyons.

The bottom line is that the brane-induced low-energy dynamics of
closed string theory in the presence of a B-field is much richer than
the one contained in the deformed action proposed in
\cite{munich1,munich2} (as, for that matter, in any other
noncommutative deformation of gravity based only on star-products). In
the following we will try to shed some light on the physical reason of
why string theory does not yield the noncommutative deformation of
gravity presented in Section \ref{NCaction}.

\section{Star-deformed symmetries versus twisted symmetries}
\label{explanation}

In the last section we have seen how string theory is unable to
account for any noncommutative theory of gravity based on
star-products, in particular the one proposed in
\cite{munich1,munich2}. In spite of all the caveats to be kept in mind
while trying to derive noncommutative gravity from the Seiberg-Witten
limit, it is quite surprising that the situation is so different from
the one arising in gauge theories, where the Seiberg-Witten limit
leads to a well-defined noncommutative gauge theory with gauge
invariance deformed appropriately.

In this section we attempt to give an explanation of why string theory
does not reproduce noncommutative gravity. In order to do that we are
going to propose an interpretation of the twisted diffeomorphisms that
allows, in our view, a better understanding of the r\^ole of this
twisted symmetry in field theory.

\subsection{Diffeomorphisms}

Let us consider an arbitrary diffeomorphism generated by a vector
field $\xi(x)=\xi^{\mu}(x)\partial_{\mu}$. We consider
two fields $\Phi_{1}$, $\Phi_{2}$ transforming under a finite
diffeomorphism generated by a vector field $\xi$ in two different
representations $\mathcal{R}_{1}$, $\mathcal{R}_{2}$
\begin{eqnarray}
\Phi_{1}'=\mathcal{D}_{1}(\xi)\Phi_{1}, \hspace*{1cm}
\Phi_{2}'=\mathcal{D}_{2}(\xi)\Phi_{2}, 
\hspace*{1cm} \mathcal{D}_{1}(\xi)\in\mathcal{R}_{1}, \hspace*{0.5cm}
\mathcal{D}_{2}(\xi)\in\mathcal{R}_{2}.
\end{eqnarray}
If the star-product of the two fields $\Phi_{1}\star\Phi_{2}$
transforms in the product representation $\mathcal{R}_{1}\otimes
\mathcal{R}_{2}$ then $\mathcal{F}^{-1}\Phi_{1}\otimes \Phi_{2}$ has to 
transform as
\begin{eqnarray}
(\mathcal{F}^{-1}\Phi_{1}\otimes\Phi_{2})'\equiv [\mathcal{D}_{1}(\xi)
\otimes\mathcal{D}_{2}(\xi)](\mathcal{F}^{-1}\Phi_{1}\otimes\Phi_{2}).
\end{eqnarray}
Inserting now the identity $\mathbf{1}=[\mathcal{D}_{1}(\xi)^{-1}
\otimes\mathcal{D}_{2}(\xi)^{-1}][\mathcal{D}_{1}(\xi)
\otimes\mathcal{D}_{2}(\xi)]$ we find
\begin{eqnarray}
(\mathcal{F}^{-1}\Phi_{1}\otimes\Phi_{2})'&=&\Big\{[\mathcal{D}_{1}(\xi)\otimes
\mathcal{D}_{2}(\xi)]\mathcal{F}^{-1}[\mathcal{D}_{1}(\xi)^{-1}
\otimes\mathcal{D}_{2}(\xi)^{-1}]\Big\}\Big\{[\mathcal{D}_{1}(\xi)
\otimes\mathcal{D}_{2}(\xi)](\Phi_{1}\otimes\Phi_{2})\Big\} \nonumber\\
&=&\mathcal{F}'{}^{-1}\Phi_{1}'\otimes\Phi_{2}',
\end{eqnarray}
where we have introduced the transformed operator 
\begin{eqnarray}
\mathcal{F}'{}^{-1}=[\mathcal{D}_{1}(\xi)\otimes
\mathcal{D}_{2}(\xi)]\mathcal{F}^{-1}[\mathcal{D}_{1}(\xi)^{-1}
\otimes\mathcal{D}_{2}(\xi)^{-1}].
\end{eqnarray}
Actually, by expanding $\mathcal{F}^{-1}$ in powers of the noncommutativity 
parameter $\theta^{\mu\nu}$ and inserting the identity, the transformed
twist operator can be written as
\begin{eqnarray}
\mathcal{F}'{}^{-1}
=e^{{i\over 2}\theta^{\mu\nu}[\mathcal{D}_{1}(\xi)\partial_{\mu}
\mathcal{D}_{1}(\xi)^{-1}]\otimes [\mathcal{D}_{2}(\xi)\partial_{\mu}
\mathcal{D}_{2}(\xi)^{-1}]}.
\end{eqnarray}

Let us now focus on infinitesimal diffeomorphisms. By writing
\begin{eqnarray}
\mathcal{D}_{1}(\xi)=e^{\delta_{\xi,1}}, \hspace*{1cm}
\mathcal{D}_{2}(\xi)=e^{\delta_{\xi,2}}
\end{eqnarray}
we find, at first order in the generators $\delta_{\xi,1}$, 
$\delta_{\xi,2}$
\begin{eqnarray}
\hspace*{-1cm}
[\mathcal{D}_{1}(\xi)\partial_{\mu}
\mathcal{D}_{1}(\xi)^{-1}]\otimes [\mathcal{D}_{2}(\xi)\partial_{\mu}
\mathcal{D}_{2}(\xi)^{-1}]&=&\partial_{\mu}\otimes \partial_{\nu}
-[\partial_{\mu},\delta_{\xi,1}]\otimes\partial_{\nu}-
\partial_{\mu}\otimes[\partial_{\nu},\delta_{\xi,2}],
\end{eqnarray}
so the variation of the twist operator $\delta_{\xi}\mathcal{F}^{-1}=
\mathcal{F}'{}^{-1}-\mathcal{F}^{-1}$ is given by
\begin{eqnarray}
\delta_{\xi}{\mathcal{F}^{-1}}=\left.
e^{{i\over 2}\theta^{\mu\nu}\partial_{\mu}\otimes\partial_{\nu}
-{i\over 2}\theta^{\mu\nu}
[\partial_{\mu},\delta_{\xi,1}]\otimes\partial_{\nu}
-{i\over 2}\partial_{\mu}\otimes[\partial_{\nu},\delta_{\xi,2}]}
\right|_{\delta},
\label{fprime}
\end{eqnarray}
where the subscript $\delta$ indicates that we should only keep the terms
linear in $\delta_{\xi,1}$ and $\delta_{\xi,2}$. 

In order to work out the expression (\ref{fprime}) we make use
of the relation
\begin{eqnarray}
\left.e^{A+\delta A}\right|_{\delta A}=
\int_{0}^{1}ds\, e^{sA}\delta A e^{(1-s)A}=
e^{A}\int_{0}^{1}dt\, e^{-tA}\delta A e^{tA},
\label{hgff}
\end{eqnarray}
which together with Hadamard's formula (\ref{hadamard})
leads to the following expression
\begin{eqnarray}
\left.e^{A+\delta A}\right|_{\delta A}=e^{A}\sum_{n=0}^{\infty}
{(-1)^{n}\over (n+1)!}[\,\underbrace{
A,[A,\ldots[A}_{n},\delta A]\ldots]],
\label{hgff2}
\end{eqnarray}
so $\delta_{\xi}\mathcal{F}^{-1}$ can be written as the following
formal series linear in $\delta_{1,\xi}$, $\delta_{2,\xi}$
\begin{eqnarray}
\delta_{\xi}\mathcal{F}^{-1}&=&
\mathcal{F}^{-1}\sum_{n=1}^{\infty}{(-i/2)^{n}\over n!}\theta^{\mu_{1}\nu_
{1}}
\theta^{\mu_{2}\nu_{2}}
\ldots\theta^{\mu_{n}\nu_{n}}\Big\{
[\partial_{\mu_{1}},[\partial_{\mu_{2}},
\ldots[\partial_{\mu_{n}},\delta_{\xi,1}]\ldots]]\otimes
\partial_{\nu_{1}}\partial_{\nu_{2}}\ldots\partial_{\nu_{n}} 
\nonumber \\
& & +\,\,\partial_{\mu_{1}}\partial_{\mu_{2}}\ldots\partial_{\mu_{n}}
\otimes
[\partial_{\nu_{1}},[\partial_{\nu_{2}},
\ldots[\partial_{\nu_{n}},\delta_{\xi,2}]\ldots]]\Big\}.
\end{eqnarray}
Therefore, the transformation of the star-product of 
$\Phi_{1}$ and $\Phi_{2}$ is given by
\begin{eqnarray}
\delta_{\xi}[\Phi_{1}(x)\star\Phi_{2}(x)]&\equiv&
\mu[\mathcal{F}'{}^{-1}\Phi_{1}'(x)\otimes \Phi_{2}'(x)]
-\mu[\mathcal{F}^{-1}\Phi_{1}(x)\otimes\Phi_{2}(x)] \nonumber \\
&=&\mu[\mathcal{F}^{-1}\delta_{\xi,1}\Phi_{1}(x)\otimes\Phi_{2}(x)]
+\mu[\mathcal{F}^{-1}\Phi_{1}(x)\otimes\delta_{\xi,2}\Phi_{2}(x)]
\nonumber \\ & &
+\,\,\mu[(\delta_{\xi}\mathcal{F}^{-1})\Phi_{1}(x)\otimes\Phi_{2}(x)]
\\ &=&
\delta_{\xi,1}\Phi_{1}(x)\star\Phi_{2}(x)+\Phi_{1}(x)\star\delta_{\xi,2}
\Phi_{2}(x)+\Phi_{1}(x)(\delta_{\xi}\star)\Phi_{2}(x),\nonumber 
\label{leibniz2}
\end{eqnarray}
where we have introduced the notation
\begin{eqnarray}
\Phi_{1}(x)(\delta_{\xi}\star)\Phi_{2}(x)\equiv \mu[(\delta_{\xi}
\mathcal{F}^{-1})
\Phi_{1}(x)\otimes\Phi_{2}(x)].
\end{eqnarray}
Using Eq. (\ref{fprime}) we can write this extra term explicitly as
\begin{eqnarray}
\Phi_{1}(x)(\delta_{\xi}\star)\Phi_{2}(x)&=&
\sum_{n=1}^{\infty}{(-i/2)^{n}\over n!}\theta^{\mu_{1}\nu_{1}}
\ldots\theta^{\mu_{n}\nu_{n}}\Big\{
[\partial_{\mu_{1}},\ldots[\partial_{\mu_{n}},\delta_{\xi,1}]\ldots]
\Phi_{1}(x)
\star\partial_{\nu_{1}}\ldots\partial_{\nu_{n}}\Phi_{2}(x) \nonumber \\
& & +\,\,\partial_{\mu_{1}}\ldots\partial_{\mu_{n}}\Phi_{1}(x)\star
[\partial_{\nu_{1}},\ldots[\partial_{\nu_{n}},\delta_{\xi,2}]\ldots]
\Phi_{2}(x)\Big\}.
\label{delta_star}
\end{eqnarray}

In this calculation we have retrieved the twisted Leibniz rule that
was obtained by twisting the Hopf algebra coproduct by the twist
operator $\mathcal{F}$. However, our analysis lead us to a different
interpretation of the twisted Leibniz rule.  Instead of thinking in
term of a twisted symmetry we can interpret the deformed
diffeomorphisms as the ordinary ones but with the additional condition
that these act not only on the fields but on the star-product as well,
according to Eq. (\ref{delta_star}). In other words, the twisted
Leibniz rule emerges from the application of the {\sl standard} one to
$\Phi_{1}\star\Phi_{2}$ and taking into account the transformation of
the star-product itself
\begin{eqnarray}
\delta_{\xi}(\Phi_{1}\star\Phi_{2})=(\delta_{\xi}\Phi)\star\Phi_{2}
+\Phi_{1}\star(\delta_{\xi}\Phi_{2})+\Phi_{1}(\delta_{\xi}\star)\Phi_
{2},
\label{extra-terms-LR}
\end{eqnarray}
i.e., in transforming a star-product of operators we have consider 
the star-product as a differential operator with its own transformation
properties. Notice that the transformation of the star-product depends
on the representation of the fields we multiply.

This interpretation of the deformed diffeomorphisms actually allows a
better understanding of the results found so far. In particular we see
that the deformed diffeomorphisms, although leaving the deformed
Einstein-Hilbert action invariant, are not {\it bona fide} physical
symmetries, since they do not act just on the fields, but on the
star-products as well. This prevents the application of the standard
Noether procedure to obtain conserved currents. The same can be said
with respect to Ward identities in the quantum theory.

It is interesting to particularize our analysis to the case of linear
affine transformations, where we can recover the results of Ref.
\cite{GBLRRV}. Considering, for simplicity, the product of two 
scalar fields $\Phi_{1}(x)$, $\Phi_{2}(x)$ the linear affine
coordinate transformation
\begin{eqnarray}
\delta x^{\mu}=B^{\mu}_{\,\,\,\,\nu}x^{\nu}+a^{\mu}.
\end{eqnarray}
induce the following tranformation for the scalar fields 
\begin{eqnarray}
\delta\Phi(x)=-B^{\mu}_{\,\,\,\,\nu}x^{\nu}\partial_{\mu}\Phi(x)-
a^{\mu}\partial_{\mu}\Phi(x)\equiv \left(-B^{\mu}_{\,\,\,\,\nu}x^{\nu}
\partial_{\mu}-a^{\mu}\partial_{\mu}\right)\Phi(x).
\end{eqnarray}
This implies that 
\begin{eqnarray}
[\partial_{\mu},\delta]=-B^{\alpha}_{\,\,\,\,\mu}\partial_{\alpha}
\end{eqnarray}
and, as a result, only the term with $n=1$ in Eq. (\ref{delta_star})
survives so we find
\begin{eqnarray}
\Phi_{1}(x)(\delta\star)\Phi_{2}(x)&=&{i\over 2}\theta^{\mu\nu}\left[
B^{\alpha}_{\,\,\,\,\mu}\partial_{\alpha}\Phi_{1}\star\partial_{\nu}
\Phi_{2}(x)+B^{\alpha}_{\,\,\,\,\nu}
\partial_{\mu}\Phi_{1}\star\partial_{\alpha}\Phi_{2}(x)\right] \nonumber\\
&=&{i\over 2}\left(B^{\alpha}_{\,\,\,\,\mu}\theta^{\mu\nu}+
\theta^{\alpha\sigma}B^{\nu}_{\,\,\,\,\sigma}\right)\partial_{\alpha}
\Phi_{1}(x)\star\partial_{\nu}\Phi_{2}(x)
\label{extra_linear}
\end{eqnarray}
The interesting thing about the case of linear affine transformation is that
this extra term in the Leibniz rule can actually be reabsorbed by a 
simultaneous transformation of the noncommutativity parameter \cite{GBLRRV}
\begin{eqnarray}
\delta\theta^{\mu\nu}=-\left(B^{\mu}_{\,\,\,\,\alpha}\theta^{\alpha\nu}+
\theta^{\mu\sigma}B^{\nu}_{\,\,\,\,\sigma}\right),
\label{theta_trans}
\end{eqnarray}
since in this case the transformation of $\Phi_{1}(x)\star_{\theta}
\Phi_{2}(x)$ picks up an extra
term associated to the transformation of $\theta^{\mu\nu}$ 
itself given by\footnote{For the sake of clarity, here and in the
remain of this subsection we have indicated by $\star_{\theta}$ the
explicit dependence of the star-product on $\theta^{\mu\nu}$.}
\begin{eqnarray}
\delta_{\theta}[\Phi_{1}(x)\star_{\theta}\Phi_{2}(x)]&\equiv& 
\Phi_{1}(x)\star_{\theta+\delta\theta}\Phi_{2}(x)-
\Phi_{1}(x)\star_{\theta}\Phi_{2}(x) \nonumber \\
&=&{i\over 2}\delta\theta^{\mu\nu}
\partial_{\mu}\Phi_{1}(x)\star_{\theta}\partial_{\nu}\Phi_{2}(x).
\label{new}
\end{eqnarray}
This cancels exactly the extra term (\ref{extra_linear}) in the
twisted Leibniz rule and one is left with 
\begin{eqnarray}
\delta_{\rm total}[\Phi_{1}(x)\star_{\theta}\Phi_{2}(x)]&\equiv &
\delta[\Phi_{1}(x)\star_{\theta}\Phi_{2}(x)]+\delta_{\theta}
[\Phi_{1}(x)\star_{\theta}\Phi_{2}(x)] \nonumber \\
&=& \delta\Phi_{1}(x)\star_{\theta}\Phi_{2}
+\Phi_{1}(x)\star_{\theta}\delta\Phi_{2}(x), 
\end{eqnarray}
where $\delta$ indicates the variation of at constant
$\theta^{\mu\nu}$ given by Eq. (\ref{extra-terms-LR}).  Hence we have
recovered the result of \cite{GBLRRV} that the star-product is
covariant under affine linear transformations provided the
noncommutativity parameter is also transformed.  In particular, if one
takes affine linear transformations belonging to the ``little group'',
i.e.  those leaving $\theta^{\mu\nu}$ invariant, the extra term in
the Leibniz rule (\ref{extra_linear}) vanishes.

It is important to keep in mind that for this to work it is crucial
that in the case of affine linear transformation the noncommutativity
parameter does not pick up any dependence on the space-time
coordinates. Of course this is not the case for general nonlinear
transformations of the coordinates, in which case there is no
transformation of the (constant) noncommutativity parameter that allows a
covariantization of the Moyal star-product.

\subsection{Gauge theories}

The situation in gauge theories is somewhat different.  It is crucial
that in this case we have in principle the Seiberg-Witten map relating
ordinary gauge transformations to their star-deformation in a well
defined series in $\theta^{\mu\nu}$, in such a way that at each stage the
effective action is invariant in the ordinary sense.  Gauge
transformations are deformed in such a way that the star-product does
not transform under a star-gauge transformation. This is clear in
the case of an adjoint field $\Phi$ on which a finite star-gauge
transformations act by
\begin{eqnarray}
\Phi(x)\longrightarrow \mathcal{U}(x)_{\star}
\star \Phi(x)\star\mathcal{U}(x)^{-1}_{\star},
\end{eqnarray}
where the inverse is defined in the sense of the star-product
\begin{eqnarray}
\mathcal{U}_{\star}(x)\star\mathcal{U}_{\star}(x)^{-1}
=\mathcal{U}_{\star}(x)^{-1}\star
\mathcal{U}_{\star}(x)=\mathbf{1}.
\end{eqnarray}
This identity guarantees the covariance of the star-product under
star-gauge transformations. Indeed, given two adjoint fields $\Phi_{1}
(x)$,
$\Phi_{2}(x)$, one finds
\begin{eqnarray}
\Phi_{i}'(x)\star\Phi_{2}'(x)&=&
[\mathcal{U}_{\star}(x)\star\Phi_{1}(x)\star\mathcal{U}_{\star}(x)^
{-1}]\star
[\mathcal{U}_{\star}(x)\star\Phi_{1}(x)\star\mathcal{U}_{\star}(x)^{-1}]
\nonumber \\
&=&\mathcal{U}_{\star}(x)\star\Phi_{1}(x)\star[\mathcal{U}_{\star}(x)^
{-1}\star
\mathcal{U}_{\star}(x)]\star\Phi_{2}\star\mathcal{U}_{\star}(x)^{-1}
\nonumber \\
&=&\mathcal{U}_{\star}(x)\star[\Phi_{1}(x)\star\Phi_{2}(x)]
\star\mathcal{U}_{\star}(x)^{-1}=[\Phi_{1}(x)\star\Phi_{2}(x)]',
\end{eqnarray}
where the prime indicates the gauge transformed field. At the 
infinitesimal
level, this means that the star-gauge variation of the
star-product of two operators $\mathcal{O}_{1}$ and $\mathcal{O}_{2}$
can be computed using the standard Leibniz rule
\begin{eqnarray}
\delta_{\rm gauge}(\mathcal{O}_{1}\star\mathcal{O}_{2})
=(\delta_{\rm gauge}\mathcal{O}_{1})\star\mathcal{O}_{2}
+\mathcal{O}_{1}\star(\delta_{\rm gauge}\mathcal{O}_{2}).
\end{eqnarray}
Using the language introduced above we can consider that the
star-product is invariant under star-gauge transformations,
i.e. $\delta_{\rm gauge}\star=0$.  A very important consequence of
this deformation of gauge invariance is that it implies a
restriction on the possible gauge groups which are reduced
to U($N$).

A second alternative \cite{munich_gauge,gauge_twist} consists of
keeping the gauge transformations undeformed
$\delta_{\omega}\Phi=iT_{\omega}\Phi$ and then twist the coproduct as
in Eq. (\ref{twisted_coproduct})
\begin{eqnarray}
\Delta(T_{\omega})_{\mathcal{F}}&\equiv& \mathcal{F}
(T_{\omega}\otimes\mathbf{1}+\mathbf{1}\otimes T_{\omega})\mathcal{F}^{-1}
=
T_{\omega}\otimes\mathbf{1}
+\mathbf{1}\otimes T_{\omega} \nonumber \\
&+&\sum_{n=1}^{\infty}{(-i/2)^{n}\over n!}\theta^{\mu_{1}\nu_{1}}
\theta^{\mu_{2}\nu_{2}}\ldots\theta^{\mu_{n}\nu_{n}}\Big\{
[\partial_{\mu_{1}},[\partial_{\mu_{2}},\ldots[\partial_{\mu_{n}},
iT_{\omega}]\ldots]]\otimes\partial_{\nu_{1}}\partial_{\nu_{2}}
\ldots\partial_{\nu_{n}} \nonumber  \\
& & +\,\,\partial_{\mu_{1}}\partial_{\mu_{2}}\ldots\partial_{\mu_{n}}
\otimes [\partial_{\nu_{1}},[\partial_{\nu_{2}},\ldots[\partial_{\nu_{n}},
iT_{\omega}]\ldots]]\Big\}.
\label{twisted_L_gauge}
\end{eqnarray}
This results in a deformation of the Leibniz rule. In the case of
adjoint fields\footnote{Now we consider the standard inverse
$\mathcal{U}(x)^{-1}\mathcal{U}(x)=\mathcal{U}(x)^{-1}\mathcal{U}(x)=
\mathbf{1}$.},
\begin{eqnarray}
\Phi(x)\longrightarrow \mathcal{U}(x)\Phi(x)\mathcal{U}(x)^{-1},
\end{eqnarray}
the extra terms in the twisted Leibniz rule reflect the fact that
\begin{eqnarray}
\Phi_{i}'(x)\star\Phi_{2}'(x)&=&
[\mathcal{U}(x)\Phi_{1}(x)\mathcal{U}(x)^{-1}]\star
[\mathcal{U}(x)\Phi_{1}(x)\mathcal{U}(x)^{-1}] \nonumber \\
&\neq& \mathcal{U}(x)[\Phi_{1}(x)\star\Phi_{2}(x)]
\mathcal{U}(x)^{-1}.
\end{eqnarray}
Interestingly, unlike the case of star-gauge transformations, there is
no restriction on the possible gauge groups that can be twisted. 

Actually we can be more general by repeating the analysis performed
above for the diffeomorphisms this time applied to gauge
transformations.  The covariance of the star-product of two fields
$\Phi_{1}$, $\Phi_{2}$ transforming respectively under finite gauge
transformations in two representations $\mathcal{R}_{1}$,
$\mathcal{R}_{2}$ as
\begin{eqnarray}
\Phi_{1}'=\mathcal{U}_{1}\Phi_{1}, 
\hspace*{1cm} \Phi_{2}'=\mathcal{U}_{2}\Phi_{2}, \hspace*{1cm}
\mathcal{U}_{1}\in\mathcal{R}_{1}, \hspace*{0.5cm}
\mathcal{U}_{2}\in\mathcal{R}_{2}
\end{eqnarray}
leads to a transformation of $\mathcal{F}^{-1}$ to
$\mathcal{F}'{}^{-1}$ given by 
\begin{eqnarray}
\mathcal{F}'{}^{-1}=e^{{i\over 2}\theta^{\mu\nu}(\mathcal{U}_{1}\partial_{\mu}
\mathcal{U}_{1}^{-1})\otimes 
(\mathcal{U}_{2}\partial_{\nu}\mathcal{U}_{2}^{-1})}.
\end{eqnarray}

Writing now $\mathcal{U}_{1}=e^{iT_{\omega}^{(1)}}$, 
$\mathcal{U}_{2}=e^{iT_{\omega}^{(2)}}$ we can write the variation 
of $\mathcal{F}^{-1}$ under a gauge transformation as
\begin{eqnarray}
\delta_{\epsilon}\mathcal{F}^{-1}&=& 
\mathcal{F}^{-1}\sum_{n=1}^{\infty}{(-i/2)^{n}\over n!}\theta^{\mu_{1}\nu_{1}}
\theta^{\mu_{2}\nu_{2}}
\ldots\theta^{\mu_{n}\nu_{n}}\Big\{
[\partial_{\mu_{1}},[\partial_{\mu_{2}},
\ldots[\partial_{\mu_{n}},T^{(1)}_{\omega}]\ldots]]\otimes
\partial_{\nu_{1}}\partial_{\nu_{2}}\ldots\partial_{\nu_{n}} 
\nonumber \\
& & +\,\,\partial_{\mu_{1}}\partial_{\mu_{2}}\ldots\partial_{\mu_{n}}
\otimes
[\partial_{\nu_{1}},[\partial_{\nu_{2}},
\ldots[\partial_{\nu_{n}},iT^{(2)}_{\omega}]\ldots]]\Big\}.
\end{eqnarray}
This expression produces again the extra terms in the twisted Leibniz
rule (\ref{twisted_L_gauge}). As in the case of diffeomorphisms
discussed above, the twisted Leibniz rule can be thought of again as
resulting from the noninvariance of the star-product
$\delta_{\omega}\star\neq 0$ under standard, i.e. non-star, gauge
transformations
\begin{eqnarray}
\delta_{\omega}(\Phi_{1}\star\Phi_{2})=[iT_{\omega}^{(1)}\Phi_{1}]\star
\Phi_{2}+\Phi_{1}\star[iT_{\omega}^{(2)}\Phi_{2}]
+\Phi_{1}(\delta_{\omega}\star)
\Phi_{2},
\end{eqnarray}
where the product $\Phi_{1}\star\Phi_{2}$ transforms now in the
product representation $\mathcal{R}_{1}\otimes\mathcal{R}_{2}$.  The
star-product is invariant only in the case of global transformations,
for which $[\partial_{\mu},iT_{\omega}]=0$.  It is in this case that
the standard Leibniz rule is retrieved.

\subsection{Discussion}

Our analysis of gauge transformations in noncommutative theories shows
that, in extending gauge symmetries to the noncommutative realm one is
faced with a choice. Either gauge transformations are deformed in such
a way that the standard Leibniz rule is satisfied or one keeps the
gauge transformations as in the commutative case at the price of
giving up the Leibniz rule.  In the latter case the new rule to
compute the gauge variation of the star-products of fields can be seen
as resulting from a twist in the Hopf algebra structure of the
universal enveloping algebra of the Lie algebra of the gauge group
extended by translations.

The obvious advantage of deforming gauge transformations into star-gauge
transformations is that gauge symmetries acts then only on the fields in
a similar way as in the commutative theories. In this sense star-gauge
symmetry is a {\it bona fide} physical symmetry, it can be implemented
in the quantum case leading to Ward identities.

On the other hand, if ordinary gauge transformations are retained
and a twisted Leibniz rule is introduced, the situation is not so
clear.  In this case the noncovariant nature of the star-product under
``undeformed'' gauge transformations has to be taken into account in
order for the action to be invariant, which amounts to replacing the
ordinary Leibniz rule with the twisted one. This means that, unlike
the previous case, now the transformations do not act only on the
fields. As a consequence this is not a physical symmetry in the usual
sense and it is not clear whether Noether charges and Ward identities
can be derived.

Let us consider the case of U($N$) noncommutative gauge
theories.  What makes these theories special is the fact that the same
action
\begin{eqnarray}
S=-{1\over 4g^{2}}\int d^{4}x \,{\rm tr\,}[F_{\mu\nu}\star F^{\mu\nu}],
\hspace*{1cm}
F_{\mu\nu}=\partial_{\mu}A_{\nu}-\partial_{\nu}A_{\mu}-i[A_{\mu},
A_{\nu}]_{\star}
\label{ncYM}
\end{eqnarray}
is invariant both under U($N$) star-gauge transformations and U($N$)
twisted gauge transformations. In Ref. \cite{munich_gauge} a  
conserved charge
was found, given by
\begin{eqnarray}
j^{\mu}=i[F^{\mu\nu},A_{\nu}]_{\star}, \hspace*{1cm}
\partial_{\mu}j^{\mu}=0.
\label{current}
\end{eqnarray}
The fact that the action (\ref{ncYM}) has both types of invariances
makes the physical interpretation of this current
unclear. Indeed, the current (\ref{current}) can be obtained from
(\ref{ncYM}) using the standard Noether procedure with respect to the
``global'' transformation $\delta
A_{\mu}=i[\omega(x),A_{\mu}]_{\star}$ and setting $\omega(x)$ to a
constant at the end of the calculation. The resulting conservation law
is consistent with star-gauge invariance. Namely,
\begin{eqnarray}
\delta_{\star\omega}j^{\mu}=i[F^{\mu\nu},\partial_{\nu}\omega]_{\star}
+[[F^{\mu\nu},A_{\nu}]_{\star},\omega]_{\star}
\end{eqnarray}
which applying the equations of motion, $\partial_{\nu}F^{\mu\nu}=-
i[F^{\mu\nu},A_{\nu}]_{\star}$, gives
\begin{eqnarray}
\delta_{\star\omega} j^{\mu}
= i\partial_{\nu}[F^{\mu\nu},\omega]_{\star}, \hspace*{1cm}
\partial_{\mu}(\delta_{\star\omega}j^{\mu})=0.
\end{eqnarray}

In the case of twisted gauge transformations, in the absence of a
Noether procedure valid for twisted symmetries, the only way to obtain
the current (\ref{current}) is as an integrability condition for the
equations of motion. In the U(1) case $j^{\mu}$ is nevertheless
invariant under twisted gauge transformations,
$\delta_{\omega}j^{\mu}=0$, while for U($N$) the
transformation of the current is given by
\begin{eqnarray}
\delta_{\omega}j^{\mu}=
i\partial_{\nu}[F^{\mu\nu},\omega]
+i[\omega,\partial_{\nu}F^{\mu\nu}+i[F^{\mu\nu},A_{\nu}]_{\star}],
\end{eqnarray}
also compatible with current conservation after applying the
equations of motion, $\partial_{\mu}(\delta_{\omega}j^{\mu})=0$.
As discussed above, the action of the twisted gauge transformations
can be seen as the action of ordinary gauge transformations acting not
only on the fields but on the star-products as well.

Because of the simultaneous presence of both types of symmetries in
the action (\ref{ncYM}) it is not easy to decide whether the origin of
the conserved current in noncommutative gauge theories is star-gauge
invariance or twisted gauge transformations. One possibility is that
in this case star-gauge invariance plays the r\^ole of a custodial
standard symmetry that forces not only that the low energy action is
expressed exclusively in terms of star-products but also the existence
of conserved currents and Ward identities. If this is the case twisted
gauge transformations might play only an accidental r\^ole in the
dynamics of noncommutative gauge theories. Of course, our discussion
only applies to U($N$) noncommutative gauge theories. In the case of
other gauge groups we are left only with twisted gauge transformations
as the invariance of the theory since star-gauge transformations 
cannot be implemented.

In the case of noncommutative gravity the apparent absence of a
star-deformed version of diffeomorphism invariance might be at the
heart of the difficulties in obtaining a noncommutative gravity action
from string theory. Comparing with the case of noncommutative gauge
theories it seems that in the case of gravity there is no symmetry of
the standard type that plays the custodial r\^ole that star-gauge
symmetry might be playing for gauge theories. Apparently twisted
symmetries by themselves are not handled by string theory so well as
standard symmetries acting only on the fields.

\section{Conclusions and outlook}
\label{conclusions}

In this paper we have studied the possibility of obtaining  
noncommutative
gravitational dynamics from string theory by studying the Seiberg-Witten
limit of the graviton interactions induced by a space-filling brane in
bosonic string theory. In particular our main interest is to investigate
whether string theory can provide some ultraviolet completion of  
recently
proposed noncommutative deformations of gravity based on the invariance
under twisted diffeomorphisms \cite{munich1,munich2}.

The conclusion of our work is that, in the case of gravitational
interactions, string theory contains much richer dynamics than those
codified in terms of star-products. We have
found that the gravitational action induced on the brane in the  
presence of
a constant $B$-field in the Seiberg-Witten limit cannot be expressed
in terms of star-products alone, unlike the action for  
noncommutative
gravity proposed in \cite{munich1,munich2}.

The consequences of this result are still to be fully understood.  In
particular it would be very interesting to clarify the r\^ole played
by twisted symmetries in the context of string theory. In the case of
noncommutative gauge theories string theory provides in the Seiberg-
Witten limit a theory which in addition to star-gauge symmetry also
has a twisted invariance. So far it has not been possible to decide
whether this twisted invariance plays a fundamental dynamical r\^ole
in string theory or whether it should be regarded as an accidental
symmetry additional to star-gauge invariance which would play a
custodial role ensuring the existence of conserved currents and 
Ward identities.

There are several ways in which one can expect to find twisted  
symmetries
in the context of open strings in the Seiberg-Witten limit. In  
particular
in this limit strings become rigid rods with variable length and their
gauge theory Fock space description is similar to that of open  
strings, in the sense
that it is the product of two copies associated with
the Hilbert space at each endpoint,
$\mathcal{H}_{1}\otimes\mathcal{H}_{2}$. Since while propagating these
rods sweep the background magnetic flux, it can be expected
that gauge transformations are twisted due to the presence of
the background field. Open string field theory in the presence of
constant $B$-fields might be specially suited to study this issue in  
detail
\cite{openSFT}.

\section*{Acknowledgments}

We would like to thank Jos\'e M. Gracia-Bond\'{\i}a, Kerstin E. Kunze,
Juan L. Ma\~nes, Rodolfo Russo and Julius Wess for interesting
discussions and comments. F.M. wants to thank the Max-Planck Institut
f\"ur Physik for financial support, and the CERN Theory Group where
part of the work reported here was done.  M.A.V.-M. acknowledges the
support from Spanish Science Ministry Grants FPA2005-04823 and
BFM2003-02121, and thanks the CERN Theory Group for hospitality and
support during the course of this work.

\section*{Appendix A. Hopf algebras: a summary of useful
formulae and definitions}
\renewcommand{\thesection}{A}

Our aim in this Appendix is to summarize basic facts about Hopf
algebras and to introduce the notation used in the paper. A more
detailed introduction to the subject of Hopf algebras and quantum
groups can be found in standard reviews (for example,
\cite{chari_pressley}).

\paragraph{Coalgebras.} The concept of a coalgebra is in a sense  
``dual''
to that of an algebra.  If an algebra $A$ is endowed with an
associative product, a coalgebra $C$ is a vector space over a field
$\mathbb{K}$ together with a coproduct $\Delta$ which is a
bilinear map
\begin{eqnarray}
\Delta:C\longrightarrow C\otimes C.
\end{eqnarray}
In general, for any element $a$ of the algebra the action of the
coproduct $\Delta$ can always be written as
\begin{eqnarray}
\Delta(a)=\sum_{i}a_{i}^{(1)}\otimes a_{i}^{(2)},
\label{coproduct}
\end{eqnarray}
where the superscript indicates the ``copy'' of $C$ in $C\otimes C$ to
which the element belongs. In addition, the coproduct is required to be
coassociative, i.e. for all $a\in C$
\begin{eqnarray}
(\Delta\otimes I)\Delta(a)=(I\otimes\Delta)\Delta(a).
\label{coassociativity}
\end{eqnarray}
In more concrete terms, this means that if $\Delta(a)$ is given by
Eq. (\ref{coproduct})
\begin{eqnarray}
\sum_{i}\Delta(a_{i}^{(1)})\otimes a_{i}^{(2)}=
\sum_{i}a_{i}^{(1)}\otimes \Delta(a_{i}^{(2)}).
\end{eqnarray}

In the same way that an algebra can contain a unit element $e$, a
coalgebra might include a counit.  This is a map
$\overline{e}:C\longrightarrow \mathbb{K}$ that satisfies
\begin{eqnarray}
(I\otimes \overline{e})\circ \Delta=I=(\overline{e}\otimes
I)\circ\Delta.
\label{counit}
\end{eqnarray}

\paragraph{Bialgebras.} A bialgebra $B$ is a vector space over a field
$\mathbb{K}$ that is at the same time an algebra and a coalgebra. In
addition the product and the coproduct must be compatible.  This means
that for all elements $a$, $b\in B$
\begin{eqnarray}
\Delta(a\cdot b)=\sum_{i}\left(a^{(1)}_{i}\cdot b^{(1)}_{i}\right) 
\otimes
\left(a^{(2)}_{i}\cdot b^{(2)}_{i}\right)=\Delta(a)\cdot \Delta(b),
\end{eqnarray}
where $\Delta(a)=\sum_{i}a^{(1)}_{i}\otimes a^{(2)}_{i}$ and
$\Delta(b)=\sum_{i}b^{(1)}_{i}\otimes b^{(2)}_{i}$.

\paragraph{Hopf algebras.} A Hopf algebra is a bialgebra $A$ together
with a linear map $S:A\longrightarrow A$ called the antipode
which for all $a\in A$ with $\Delta(a)=\sum_{i}a^{(1)}_{i}\otimes
a^{(2)}_{i}$ satisfies
\begin{eqnarray}
\sum_{i}a_{i}^{(1)}\cdot S\left(a_{i}^{(2)}\right)=
\sum_{i}S\left(a_{i}^{(1)}\right)\cdot a_{i}^{(2)}=(e\circ\overline 
{e})(a).
\label{antipode}
\end{eqnarray}

\paragraph{Examples.}

The simplest one
is the Hopf algebra associated with a group $\mathcal{G}$. Given this
group we can always define the group algebra
$\mathbb{K}\mathcal{G}$ over a field $\mathbb{K}$ as the algebra of
lineal combinations of elements of $\mathcal{G}$ with coefficients
$\lambda\in \mathbb{K}$ and with the product
\begin{eqnarray}
(\lambda_{1}g_{1})\cdot
(\lambda_{2}g_{2})=(\lambda_{1}\lambda_{2})(g_{1}g_{2}), \hspace*{1cm}
\forall \lambda_{1},\lambda_{2}\in \mathbb{K} \,\,\,\,
g_{1},g_{2}\in \mathcal{G},
\end{eqnarray}
where on the right hand side of the equation we use the product of
$\mathbb{K}$ and $\mathcal{G}$. This algebra has the unit element
$e=I$, where $I$ the identity element of $\mathcal{G}$.

Actually, the algebra $\mathbb{K}\mathcal{G}$ has also a coalgebra
structure given by the coproduct defined by
\begin{eqnarray}
\Delta(g)=g\otimes g, \hspace*{1cm} g\in\mathcal{G}.
\label{copG}
\end{eqnarray}
It is straightforward to show that this coproduct is coassociative.
In addition the counit $\overline{e}$ is defined by $\overline{e}(g)=1$
where $1$ is the identity of the field $\mathbb{K}$. Bilinearity of
the coproduct and the linearity of the counit determines the map for
any element of $\mathbb{K}\mathcal{G}$. Moreover, the coproduct (\ref 
{copG}) is
actually compatible with the product, so $\mathbb{K}\mathcal{G}$ is  
in fact a bialgebra.
This structure can be extended to a Hopf algebra by the antipode map  
$S$ defined by
\begin{eqnarray}
S(g)=g^{-1}, \hspace*{1cm} \forall g\in\mathcal{G},
\label{antG}
\end{eqnarray}
which satisfies indeed the property (\ref{antipode}).

The second instance of Hopf algebras that we are going to study is the
universal enveloping algebra of a Lie algebra. Let $\mathcal{L}$ be a
Lie algebra over a field $\mathbb{K}$ with generators $\xi_{i}$
($i=1,\ldots,{\rm dim\,}\mathcal{L}$). The universal enveloping
algebra $\mathcal{U}(\mathcal{L})$ associated with $\mathcal{L}$ is
the algebra generated by $\xi_{i}$ with the identification
\begin{eqnarray}
\xi_{i}\cdot \xi_{j}-\xi_{j}\cdot \xi_{i}\sim [\xi_{i},\xi_{j}],  
\hspace*{1cm}
i,j=1,\ldots,{\rm dim\,}\mathcal{L},
\label{Lalgprod}
\end{eqnarray}
where $[a,b]$ denotes the commutator operation in the Lie algebra.

Given $\xi\in \mathcal{U}(\mathcal{L})$ the mapping
\begin{eqnarray}
\Delta(\xi)= \left\{
\begin{array}{cll}
\xi\otimes e+e\otimes \xi & \hspace*{0.1cm} & \xi\neq e \\
e\otimes e & \hspace*{0.1cm} & \xi=e
\end{array}
\right.
\label{coproductUEA}
\end{eqnarray}
defines a coassociative coproduct compatible with the algebra
product. A counit $\overline{e}$ is defined by
\begin{eqnarray}
\overline{e}(\xi)=
\left\{
\begin{array}{cll}
0 & \hspace*{0.1cm} & \xi\neq e \\
1 & \hspace*{0.1cm} & \xi=e
\end{array}
\right..
\end{eqnarray}
This shows that $\mathcal{U}(\mathcal{L})$ is endowed with a  
bialgebra
structure. This is actually extended to a Hopf algebra by the
antipode map
\begin{eqnarray}
S(\xi)=\left\{
\begin{array}{ccl}
-\xi &\hspace*{0.1cm} &  \xi\neq e \\
e &\hspace*{0.1cm}  & \xi=e
\end{array}
\right.,
\end{eqnarray}
which can be easily seen to satisfy the condition (\ref{antipode}).

\paragraph{The action of a Hopf algebra on an algebra.}

For the applications of Hopf algebras as twisted symmetries we need to
define its action on an algebra $A$. Roughly speaking we want to
define a map $\alpha:H\otimes A\longrightarrow A$
with the property that for every $\xi,\zeta\in H$, $a\in A$
\begin{eqnarray}
\alpha(\xi\cdot\zeta\otimes a)=\alpha[\xi\otimes \alpha(\zeta\otimes  
a)],
\hspace*{1cm} \alpha(e\otimes a)=a,
\end{eqnarray}
where $e\in H$ is the unit element.

If $H$ was just an algebra this would be the end of the
story. However $H$ has a Hopf algebra structure as well, so additional
conditions are imposed involving the coproduct and the
counit. If we denote now by $\mu$ the product map on the algebra $A$,
$\mu(a\otimes b)=ab$ for $a,b\in A$ these conditions are
\begin{eqnarray}
\alpha(\xi\otimes ab)=\mu\circ\alpha[\Delta(\xi)\otimes (a\otimes b)],
\hspace*{1cm}
\alpha(\xi\otimes \mathbf{1})=\mathbf{1}\circ \overline{e}(\xi),
\label{compatibility_action}
\end{eqnarray}
with $\mathbf{1}$ the unity element of the algebra $A$ and the action
of the map $\alpha$ is extended to
\begin{eqnarray}
\alpha[(\xi\otimes\zeta)\otimes (a\otimes b)]=
\alpha(\xi\otimes a)\otimes \alpha(\zeta\otimes b).
\end{eqnarray}

As an example we can consider the action of the universal enveloping
algebra of a Lie algebra, $\mathcal{U}(\mathcal{L})$ on an algebra
$A$. Since the coproduct $\Delta$ is given by
Eq. (\ref{coproductUEA}), we have for $a,b\in A$, $\xi\in
\mathcal{U}(\xi)$
\begin{eqnarray}
\alpha[\Delta(\xi)\otimes(a\otimes b)]&=&
\alpha(\xi\otimes a)\otimes b +
a \otimes \alpha(\xi\otimes b). \nonumber
\end{eqnarray}
Therefore Eq. (\ref{compatibility_action}) reads
\begin{eqnarray}
\alpha(\xi\otimes ab)=\mu\circ[\alpha(\xi\otimes a)\otimes b]
+\mu\circ[a\otimes \alpha(\xi\otimes b)].
\label{leibniz1}
\end{eqnarray}
This identity is nothing but Leibniz' rule giving the action of an
element of the Hopf algebra $\xi$ on the product of two algebra
elements $a$ and $b$.

\paragraph{Twisting.}
Of course the coproduct given in Eq. (\ref{coproductUEA}) is just one
among all the possible choices for a coproduct satisfying the
appropriate conditions. Other possible coproduct can be obtained from
this one by {\it twisting}. This means that given a invertible element
$\mathcal{F} \in
\mathcal{U}(\mathcal{L})\otimes\mathcal{U}(\mathcal{L})$, a new Hopf
algebra structure can be defined using the twisted coproduct
\begin{eqnarray}
\Delta_{\mathcal{F}}(\xi)\equiv
\mathcal{F}\Delta(\xi)\mathcal{F}^{-1}=
\mathcal{F}(\xi\otimes e+e\otimes\xi)\mathcal{F}^{-1},
\end{eqnarray}
and the twisted antipode
\begin{eqnarray}
S_{\mathcal{F}}(\xi)\equiv u S(\xi) u^{-1},
\end{eqnarray}
where $u$ is defined by
\begin{eqnarray}
u\equiv \mu[(I\otimes S)\mathcal{F}]=\sum_{i}f_{i}^{(1)}S(f^{(2)}_{i}).
\end{eqnarray}
In writing this last expression we have used the decomposition
\begin{eqnarray}
\mathcal{F}=\sum_{i} f_{i}^{(1)}\otimes f_{i}^{(2)}.
\end{eqnarray}
Of course not every twist operator $\mathcal{F}$ gives rise to a well
defined twisted Hopf algebra. In particular one should make sure that
the new coproduct $\Delta_{\mathcal{F}}$ preserves coasociativity
(\ref{coassociativity}) and is compatible with the counit
$\overline{e}$ (\ref{counit}). This respectively is guaranteed 
if $\mathcal{F}$ satisfies the following conditions
\begin{eqnarray}
(e\otimes \mathcal{F})(I\otimes \Delta)\mathcal{F}&=&
(\mathcal{F}\otimes e)(\Delta\otimes I)\mathcal{F} \nonumber \\
(I\otimes \overline{e})\mathcal{F}&=&(\overline{e}\otimes I)\mathcal{F}
=e.
\end{eqnarray}

\end{document}